\documentclass[twoside]{elsart}
\usepackage{amsmath,latexsym,amssymb}
\journal{Annals of Physics}

\numberwithin{equation}{section}
\newcommand{\R}{\mathbb R}
\newcommand{\C}{\mathbb C}

\newcommand{\K}{\Bold{K}}

\newcommand{\p}{\partial}

\newcommand{\Xa}{\mathbf{X}_{(\alpha)}}
\newcommand{\Xb}{\mathbf{X}_{(\beta)}}

\newcommand{\Za}{{Z^d_a}}

\newcommand{\Ta}{T^+_a}
\newcommand{\zTa}{Z^+_a}
\newcommand{\T}{{T^+_a}'}

\newcommand{\PM}{P_a^{\mathbb{V}}\mathbb{M}}

\newcommand{\DQ}[1]{\mathcal{D}_{{Q},{W}}#1}
\newcommand{\DO}[1]{\mathcal{D}#1}
\newcommand{\Da}[1]{\mathcal{D_{\nu}}#1}
\newcommand{\Bold}[1]{\mbox{\boldmath$\mathit{#1}$}}
\newcommand{\dual}{\langle \tau',\tau\rangle}
\newcommand{\zdual}{\langle z',z\rangle}

\newcommand{\xdual}{\langle x',x\rangle}

\newcommand{\taurange}{\int_{C_+}}
\newcommand{\ti}{\mathrm{t}}
\newcommand{\x}{\mathrm{x}}

\newcommand{\M}{\mathbb{M}}
\newcommand{\U}{\mathbb{U}}
\newcommand{\I}{\mathbb{I}}
\newcommand{\V}{\mathbb{V}}
\newcommand{\QED}{\hspace{.2in}\square\newline}

\newcommand{\taubot}{\tau_{\mathrm{x}_a}^\bot}

\newcommand{\tr}{\tau}

\newtheorem{theorem}{Theorem}[section]
\newtheorem{corollary}{Corollary}[section]
\newtheorem{proposition}{Proposition}[section]
\newtheorem{definition}{Definition}[section]
\newtheorem{assumption}{Restriction}[section]
\newtheorem{lemma}{Lemma}[section]
\begin{document}

\begin{frontmatter}
\title{Path Integral Solution of Linear Second Order
Partial Differential Equations I. The General Construction}
\author{J. LaChapelle}
\ead{jlachapelle@comcast.net}

\begin{abstract}
A path integral is presented that solves a general class of linear
second order partial differential equations with Dirichlet/Neumann
boundary conditions. Elementary kernels are constructed for both
Dirichlet and Neumann boundary conditions. The general solution
can be specialized to solve elliptic, parabolic, and hyperbolic
partial differential equations with boundary conditions. This
extends the well-known path integral solution of the
Schr\"{o}dinger/diffusion equation in unbounded space. The
construction is based on a framework for functional integration
introduced by Cartier/DeWitt-Morette.
\end{abstract}

\end{frontmatter}


\section{Introduction}
Functional integral solutions of second order elliptic and
parabolic partial differential equations (PDEs) have been known
for a long time. In particular, stochastic methods have been used
to solve inhomogeneous elliptic and diffusion-type parabolic PDEs
(see e.g. \cite{DY/YU}, \cite{FR}, \cite{FRI1}), and path
integrals\footnote{Although path integrals are functional
integrals, I will continue to use the former term to conform to
physics usage and to emphasize that the functions used in
stochastic methods are continuous while those used in path
integrals are $L^{2,1}$.} in physics have been used to solve the
Schr\"{o}dinger equation (\cite{FE/HI},\cite{CA/D-M}) and
homogeneous Dirichlet problem (\cite{LA}).

Stochastic methods have perhaps been the most fruitful in terms of
solving PDEs. The majority of stochastic results have been
obtained for scalar functions on $\R^n$; although Elworthy
(\cite{EL}) was able to extend some of these to tensor fields on
manifolds. However, stochastic methods are limited in scope
because they are based on the Wiener process. Consequently, they
cannot be applied to Schr\"{o}dinger-type parabolic PDEs (except
through analytic continuation), and they have nothing to say about
hyperbolic PDEs.

On the other hand, path integrals have been developed extensively
in a myriad of physics applications. They have been extended to
tensors on manifolds (following Elworthy), Grassman variables, and
fields. However, for the most part, path integrals have not been
exploited for solving PDEs outside of Schr\"{o}dinger-type
parabolic equations in unbounded space.

In this paper, a general path integral is presented that solves a
wide class of linear second order PDEs with given Dirchlet/Neumann
boundary conditions. Relevant elementary kernels are also
constructed to facilitate incorporation of Dirchlet/Neumann
boundary conditions. In a subsequent paper, the general path
integral will be specialized to solve elliptic, parabolic, and
hyperbolic PDEs; several examples will be worked out to check the
validity of the constructed solutions against known solutions; and
some new calculational techniques will be introduced.

Section \ref{sec. path integrals} contains the outline of a
framework for functional integration developed by
Cartier/DeWitt-Morette in \cite{CA/D-M} (see also \cite{CA/D-W2}).
It allows one to define path integrals in a general setting. The
three main ingredients of the framework are: an infinite
dimensional space of pointed paths which take their values in a
(complex) manifold; a parametrization of this infinite dimensional
space of pointed paths by a Banach space; and integrators (that
assume the role of problematic measures in infinite dimensional
spaces) defined on the parametrizing Banach space. The scheme is
then amended by restricting the space of integrable functionals;
and a new integrator, the gamma integrator, is introduced. These
additions play a key role in the construction of the solution of
the general PDE. Specializing various components of the functional
integration framework leads to path integrals that are solutions
of the three classes of PDEs mentioned above.

Part of the utility of the Cartier/DeWitt-Morette scheme is that
it deals with integrators instead of measures. For example, the
integrator corresponding to the Weiner measure can be used to
solve diffusion-type parabolic PDEs. However, while it is not
possible to define a \emph{measure} that corresponds to the
complex version of a Weiner measure, it is possible to define an
\emph{integrator} that does. Consequently, in this scheme, one can
solve Schr\"{o}dinger-type parabolic PDEs directly.

Since the Cartier/DeWitt-Morette formulation includes
parametrizations that are similar to stochastic differential
equations and integrators that are similar to Weiner measures, it
is not surprising that some of the solutions constructed here are
similar to the stochastic constructions. Some of the similarity
arises from the use of a dependent variable that reparametrizes
the time along a path. From a physics perspective, the variable is
a non-dynamical degree of freedom that is introduced because the
paths of interest are point-to-boundary instead of
point-to-point---the presence of the boundary induces a
constraint. Of particular importance is the minimum time to reach
a given boundary starting from some point. Alternatively, one can
consider the path-dependent time to reach a given boundary
starting from some point. In stochastic parlance this variable is
the first exit time from the boundary (the first time at which a
given stochastic path intersects the boundary). Because these two
notions are similar, they lead to similar constructions. However,
there is a crucial difference that will become clear later, and
the construction based on the former notion is advantageous in
terms of evaluating path integrals.

Section \ref{sec. main} constructs the path integral solution
along with its associated kernels for various boundary conditions.
The gamma and Gaussian integrators play a major role in the
construction. The reader should be warned that no attempt is made
here to determine conditions for existence and uniqueness of the
constructed solutions. However, it is reasonable to expect that
the path integral setting will be a valuable tool in addressing
these difficult issues, because it shifts the focus to the
function space instead of the target manifold.

One of the virtues of path integrals is that they automatically
incorporate boundary conditions; sometimes implicitly, sometimes
explicitly. As such, they offer an alternate approach to study and
solve PDEs with non-trivial boundary conditions. In particular,
complicated geometries are no different from symmetrical
geometries in principle. Since computational and approximation
methods of path integrals are fairly well developed, finding
solutions for a given boundary problem should be at least
systematic if not simplified.

Supporting details of the functional integration framework are
presented in Appendix A and Appendix B. Appendix A contains
several properties of functional integrals that are fairly
straightforward but many of which have not appeared explicitly in
the references. Appendix B discusses the well-known Gaussian
integrator and introduces the new gamma and Hermite integrators.
The Hermite integrator is a generalization of the Gaussian
integrator and the gamma integrator is the scale invariant
analogue of the (translation invariant) Gaussian integrator.

\section{Path Integration}\label{sec. path integrals}
\subsection{Cartier/DeWitt-Morette scheme}
According to the general scheme (\cite{CA/D-M}, \cite{CA/D-W2}), a
path integral is defined on a separable Banach space $X$ with a
norm $\|x\|$ where $x\in X$ is a map $x:\Sigma\rightarrow\M$.
$\Sigma$ is a $1$-dimensional manifold and $\M$ is an
$m$-dimensional manifold. The dual Banach space $X'\ni x'$ is a
space of linear forms such that $\xdual\in\C$ with an induced norm
given by
\begin{equation}
  \|x'\|=\sup_{x\neq 0}|\xdual|/\|x\|\;.\notag
\end{equation}
Assuming $X'$ is separable renders it a Polish space, and,
therefore, it admits complex Borel measures $\mu$.

Given two continuous, bounded, and $\mu$-integrable functionals
$\Theta:X\times X'\rightarrow \C^d$ and $Z:X'\rightarrow\C^d$, an
integrator on $X$ is defined by
\begin{definition}\label{def1}
\begin{equation}\label{integrator}
  \int_{X}\Theta(x,x')\,\mathcal{D}_{\Theta,Z}x:=Z(x')\;.
\end{equation}
\end{definition}

A space, $\mathcal{F}(X)$, of functionals integrable with respect
to this integrator consists of functionals defined by
\begin{definition}\label{def2}
\begin{equation}\label{integrable functions}
F_{\mu}(x):=\int_{X'}\Theta(x,x')\,d\mu(x')\;.
\end{equation}
\end{definition}
In this paper, $F_{\mu}(x)$ for some path $x\in X$ will be a
tensor field or tensor distribution of type $(r,s)$ along
$x(\Sigma)$ on the manifold $\M$. The tensor distributions
(suitably restricted) will be used to construct kernels of second
order linear partial differential operators. In the sequel, all
tensor fields and distributions are assumed to be well defined,
i.e. they are continuous, differentiable, integrable, bounded,
etc. as required for any given case. Since existence and
uniqueness issues are not addressed, this simplification is
warranted.

The map $\mu\mapsto F_{\mu}$ is required to be injective rendering
$\mathcal{F}(X)$ a Banach space, and $\mathcal{F}(X)$ can be
endowed with a norm $\|F_{\mu}\|$ defined to be the total
variation of $\mu$. An integral operator
$\int_X\mathcal{D}_{\Theta,Z}x$ on the normed Banach space
$\mathcal{F}(X)$ is then defined by
\begin{definition}\label{def3}
\begin{equation}\label{integral definition}
  \int_{X}F_{\mu}(x)\,\mathcal{D}_{\Theta,Z}x:=\int_{X'}Z(x')\,d\mu(x')\;.
\end{equation}
\end{definition}
The integral operator $\int_X\mathcal{D}_{\Theta,Z}x$ is a bounded
linear form on $\mathcal{F}(X)$ with bound $\left|\int_X
F_{\mu}\,\mathcal{D}_{\Theta,Z}x\right|\leq\|F_{\mu}\|$
(\cite{CA/D-W2}).

A fruitful generalization is to consider functionals $\Theta$ and
$Z$ that depend on a parameter $\lambda$ where, for example,
$\lambda$ may be in $\R$ or $\C$. It will be assumed that
$F_{\mu}(x;\lambda)$ is differentiable with respect to $\lambda$,
and that
$\partial_{\lambda}^{(n)}F_{\mu}(x;\lambda)\in\mathcal{F}(X)$ for
$n\in\{1,2\}$.

 The right-hand side of (\ref{integral definition}) should not be
thought of as a prescription for calculating the left-hand
side---in practice, one neither knows nor specifies $\mu$.
Instead, one invariably makes use of some form of localization to
reduce the left-hand side to a finite dimensional integral. Hence,
a `good' (or at least useful) characterization encoded in
Definition \ref{def1} must reduce to the correct finite integral
for any dimension: consequently, it is reasonable to expect path
integrals to possess properties analogous to finite dimensional
integrals. Some properties of path integrals that are particularly
relevant are listed in Appendix \ref{app. properties}.

For actual applications, the goal is to define an integral over an
infinite dimensional space, $\PM$, composed of \emph{pointed}
$L^{2,1}$ maps $x:\Sigma\rightarrow \M$ with a mutual fixed
end-point $\x_a\in \M$. Here, $\M$ is a (possibly complex)
$m$-dimensional paracompact differentiable manifold, and I will
restrict to $\Sigma=\I\subseteq\R$ or $\Sigma=S^1$ and refer to
$x\in\PM$ as a path (hence the term `path' integral).
Additionally, assume a set of $d\leq m$ linearly independent
vector fields $\Xa$ where $\alpha \in \{1,\ldots ,d\}$. The set
$\{\Xa\}$ generates a sub-bundle $\V\subseteq T\M$.

However, in general $\PM$ is not a Banach space. Analogous to the
definition of integration on finite dimensional manifolds in terms
of integration on $\R^n$, path integrals on $\PM$ can be defined
in terms of path integrals on a separable Banach space. The
definition relies on a parametrization of the paths effected by
the differential system
\begin{equation}\label{parametrization}
  \left\{ \begin{array}{ll}
    \Bold{dx}(\ti,z)-\mathbf{Y}(x(\ti,z))d\ti=
    {\Xa}(x(\ti,z))\Bold{dz}^i(\ti) \\
    x(\ti_a)=\x_a
  \end{array}
  \right.
\end{equation}
where $Z\ni z$ is a Banach space, and
$\dot{\Bold{x}}(\ti)-\mathbf{Y}(x(\ti,z))\in \V_{x(\ti)}\subseteq
T_{x(\ti)}\M$ for each $\ti\in\Sigma$ and some given vector field
$\mathbf{Y}$ on $\M$. The solution of (\ref{parametrization}) will
be denoted by $x(\ti,z)=\x_a\cdot \mathit{\Sigma}(\ti,z)$ where
$\mathit{\Sigma}(\ti,z):\M\rightarrow \M$ is a global
transformation on $\M$ such that $x(\ti_a,\cdot)=\x_a\cdot
\mathit{\Sigma}(\ti_a,\cdot)=\x_a$.\footnote{Compactness of the
manifold $\M$ may be required in some cases to ensure global
existence and uniqueness of the map $z\mapsto x$. However, one is
usually interested in a submanifold $\mathbb{U}\subseteq \M$ so
compactness of the entire manifold may not be necessary.}

The nature of the integrator on $Z$ dictates the range of the
paths $z\in Z$ (or vise-versa). For example, for a translation
invariant integrator, $z\in\mathit{P}^{\V}_0{\R}^d(\C^d)=:\Za$;
and for a scale invariant integrator,
$z\in\mathit{P}^{\V}_0{\R}_+(\C_+)=\ln(Z_a^1)=:\zTa$ with
$z(\ti_a)=0$ for some $\ti_a\in\Sigma$.\footnote{The notation
$\mathit{P}^{\V}_0\R^d(\C^d)$ is short-hand for
$\mathit{P}^{\V}_0\R^d$ or
$\mathit{P}^{\V}_0\C^d$.}$^,$\footnote{$\C_+$ denotes the right
complex plane.}$^,$\footnote{Translation and scale invariant
integrators are characterized in Proposition \ref{prop. change of
variable}.} The differential system (\ref{parametrization})
associates a pointed path $z\in \Za(\zTa)$ with each pointed path
$x\in \PM$ yielding a parametrization $P:\Za(\zTa) \rightarrow
\PM$.

The parametrization $P:\Za(\zTa) \rightarrow \PM$ and Definition
\ref{def3} allow for a rigorous definition of the path integral of
a functional $F(x;\ti)$ over $\PM$:
\begin{definition}\label{def4}
\begin{equation}\label{PM integral def.}
  \int_{\PM}F(x;\ti)\, \mathcal{D}x :=
   \int_{\Za(\zTa)}F_{\mu}(\x_a\cdot \mathit{\Sigma}(\ti,z))\,
  \mathcal{D}_{\Theta,Z}z.
\end{equation}
\end{definition}

\subsection{Restrictions}
In order to implement the analogues of invariant measures and
integration by parts in finite dimensions, it is useful to impose
two restrictions on the general scheme outlined in the previous
subsection.

Let $Y$ be a separable Banach space and $M:X\rightarrow Y$ be a
diffeomorphism with derivative mapping $M'_{(x)}:T_xX\rightarrow
T_yY$. The derivative map possesses a transpose
$\widetilde{M'}:TY'\rightarrow TX'$ defined by
\begin{equation}
  \langle \widetilde{M'}_{(y')}\Bold{y}',\Bold{x}\rangle_{T_{x}X}
  :=\langle
  \Bold{y}',M'_{(x)}\Bold{x}\rangle_{T_yY}\;.
\end{equation}
$M$ induces a germ $R:Y'\rightarrow X'$ along with its associated
derivative mapping $R'_{(y')}:T_{y'}Y'\rightarrow T_{x'}X'$ and
transpose $\widetilde{R'}_{(x)}:T_xX\rightarrow T_yY$ from the
relation
\begin{equation}
  \langle R'_{(y')}\Bold{y}',\Bold{x}\rangle_{T_xX}=\langle
  \Bold{y}',\widetilde{R}'_{(x)}\Bold{x}\rangle_{T_yY}\;.
\end{equation}
Evidently for $y=M(x)$ and $x'=R(y')$, $\widetilde{M}'=R'$ and
$\widetilde{R}'=M'$.

Define the $\mu$-integrable functionals $\overline{\Theta}:Y\times
Y'\rightarrow\C^{(r+s)}$ and $\overline{Z}:Y'\rightarrow
\C^{(r+s)}$ by
\begin{equation}\label{theta bar}
  \overline{\Theta}\circ (M\times R^{-1}):=\Theta
\end{equation}
and
\begin{equation}\label{Z bar}
  \overline{Z}\circ R^{-1}:=Z
\end{equation}
along with their associated integrator
\begin{equation}
  \int_{Y}\overline{\Theta}(y,y')
  \,\mathcal{D}_{\overline{\Theta},\overline{Z}}y
  :=\overline{Z}(y')\;.
\end{equation}

Let $R(Y')=Y'$. The first restriction relates $Z$ and
$\overline{Z}$ at the same point in $Y'$;
\begin{assumption}\label{theta relation}
\begin{equation}\label{theta relation eq.}
  \overline{Z}(y')=|\mathrm{Det}R'_{(y')}|^{-1}Z(y')
\end{equation}
\end{assumption}
for $R$ a diffeomorphism and non-vanishing
$|\mathrm{Det}R'_{(y')}|$. For the determinant to be well defined,
require that $R'$ be nuclear (see for example \cite{CA2} for
relevant details).

To state the second restriction, it is useful to interpret
$F_{\mu}(x)\,\mathcal{D}_{\Theta,Z}x=:F_{\mu}$ as a form on $X$.
Let $\mathcal{F}^{\wedge}(X)$ denote the space of integrable forms
$F_{\mu}$ on $X$. Let $\Bold{d}_y$ denote the Gateaux derivative
of $F_{\mu}(x)$ in the $y$ direction, i.e.,
\begin{equation}
  \Bold{d}_yF_{\mu}(x)
  =\lim_{h\rightarrow 0}\frac{F_{\mu}(x+hy)-F_{\mu}(x)}{h}
=\int_{\Sigma}\frac{\delta F_{\mu}(x)}{\delta
x(\ti)}\,y(\ti)\,d\ti\;.
\end{equation}
Assume $F_{\mu}(x)$ and $\mathcal{D}_{\Theta,Z}x$ are Gateaux
differentiable, then require
\begin{assumption}\label{closed form}
\begin{equation}\label{closed form eq.}
  \int_X\Bold{d}_yF_{\mu}=
  \int_X\Bold{d}_y[F_{\mu}(x)\;\mathcal{D}_{\Theta,Z}x]=0
\end{equation}
\end{assumption}
for arbitrary $y\in X$.

These two restrictions have seemingly been ``pulled out of a
hat'', but, as shown in Appendix \ref{app. properties}, they
enable the characterization of invariant integrators and
integration by parts. Henceforth, $\mathcal{F}_R(X)$ will denote
the space of integrable functionals restricted by (\ref{theta
relation eq.}) and (\ref{closed form eq.}). Likewise,
$\mathcal{F}_R^{\wedge}(X)$ is assumed to be similarly restricted.
Note that (\ref{theta relation eq.}) and (\ref{closed form eq.})
may be quite restrictive and $\mathcal{F}_R(X)$ may be severely
limited or even empty.\footnote{Clearly, the question of existence
of PDE solutions constructed in the next section is directly
related to the nature of $\mathcal{F}_R(X)$.}

\section{Path Integral Solution of PDEs}\label{sec. main}
The path integral framework introduced in Section \ref{sec. path
integrals} allows the construction of solutions to general classes
of linear second order partial differential equations. It turns
out (as shown in a subsequent paper) that various choices of the
manifold $\M$ and the parametrization $P$ will yield elliptic,
parabolic, and hyperbolic PDEs. It is noteworthy that the
construction can cover all three cases.

\subsection{General Solution}

It is now possible to construct a path integral for PDEs with
finite boundary. The construction depends crucially on a time
reparametrization $\tau$ and its associated integrator
$\DO{\tau}$. Appendix \ref{sec. variation} presents the motivation
for introducing a time reparametrization for bounded regions. It
should be pointed out that the reparametrization used here is
fundamentally different from the time reparametrization sometimes
employed in stochastic methods (see e.g. \cite{YO/D-W}) or by Duru
and Kleinert (\cite{DU/KL}): there the reparametrization is path
\emph{dependent}. Specifically, stochastic constructs parametrize
the first exit time of a given path---since the paths depend on a
random variable, then so does the first exit time. Here the
reparametrization is path \emph{independent} and the first exit
time used in this construction is for a critical path. This
difference allows $z$ and $\tr$ to be integrated independently.

With these issues in mind and some reflection on existing
functional and path integral solutions to selected PDEs, a path
integral solution of a linear second order inhomogeneous PDE is
constructed.

First some definitions:
\begin{definition}
$\Ta$ is the space of $L^{2,1}$ functions
\emph{$\tr:\mathbb{T}=[\ti_a,\ti_b]
\rightarrow[\tau_a,\tau_b]\cup[\tau_a,\tau_b]^*\subset\C_+$} such
that \emph{$|d\tr/d\ti|>0$} and \emph{$\tau_a:=\tr(\ti_a)=0$}.

$\PM$ is the space of $L^{2,1}$ pointed paths
$x:\tr(\mathbb{T})\rightarrow \M$ with $x(\tau_a)=\x_a\in
\mathbb{U}\subseteq\M$ whose velocity vectors are elements of
$\mathbb{V}\subseteq T\M$.

$\Za$ is the space of $L^{2,1}$ pointed paths
$z:\tr(\mathbb{T})\rightarrow \R^d(\C^d)$ with $z(\tau_a)=0$.
\end{definition}

\begin{definition}
 Let $x(\tr(\ti),z)=\x_a\cdot\mathit{\Sigma}(\tau(\ti),z)$---where
  $\mathit{\Sigma}(\tau(\ti),z):\M\rightarrow\M$
 is a global transformation on $\M$ such that
 $x(\tr(\ti_a),z)=\x_a\cdot\mathit{\Sigma}(\tau(\ti_a),z)=\x_a$---denote
 the solution of the parametrization $P:\Za\rightarrow\PM$
 according to
\emph{\begin{equation}\label{tau parametrization}
  \left\{ \begin{array}{ll}
    \mbox{\boldmath{$dx$}}(\tr(\ti),z)-\mathbf{Y}(x(\tr(\ti),z))d\tr=
    {\Xa}(x(\tr(\ti),z))\mbox{\boldmath{$dz$}}^\alpha(\tr(\ti)) \\
    x(\tau_a)=\x_a\;\;.
  \end{array}
  \right.
\end{equation}}
\end{definition}

\begin{definition}
The functional $\mathcal{S}(x(\tau_{b},z))$ is defined by
\begin{equation}\label{action}
  \mathcal{S}(x(\tau_{b},z)):=\pi Q(x(\tau_{b},z))
  -\int_{0}^{\tau_{b}}V(x(\ti,z))\;d\ti
  \end{equation}
where
\begin{eqnarray}
  Q(x(\tau_{b},z))
  :=\langle Dx,x\rangle
  =\int_{0}^{\tau_{b}}\int_{0}^{\tau_{b}}
  {x^i(\ti,z)}D_{ij}(\ti,\mathrm{u}){x^j(\mathrm{u},z)}
  \;d\ti\,d\mathrm{u}
\end{eqnarray}
in a local chart.
\end{definition}
The operator matrix $D_{ij}(\ti,\mathrm{u})$ with
$i,j\in\{1,\ldots,n\}$ depends on the nature of $\M$. In physical
applications, $D_{ij}$ is usually related to a metric or
symplectic form. For example, $Q$ is often identified with an
action functional of the form
\begin{equation}
  A(x)=\int_{\ti_a}^{\ti_b} h_{x(\ti)}
  \left(\dot{\Bold{x}}(\ti)-\mathbf{Y}(x(\ti))
  ,\dot{\Bold{x}}(\ti)-\mathbf{Y}(x(\ti))\right)\,d\ti
\end{equation}
where $h_\x$ is a quadratic form on $T_\x\M$ and $\mathbf{Y}\in
T\M$ is some vector field. The end-point value of the Green's
function of $D$ is the matrix
$G^{\alpha\beta}:=G^{\alpha\beta}(\ti_b,\ti_b)$ discussed in
Appendix \ref{app. gaussian integrators}.

\begin{definition}
Let $\taubot$ denote the first exit time of a \underline{critical}
path, with respect to $\mathcal{S}(\x(\tau_{b},z))$, starting at a
point $\x_a\in \mathbb{U}\subseteq\M$; i.e.,
\emph{$x_{cr}(\taubot)=:\x_a^\bot\in\partial \mathbb{U}$} and
\emph{$x_{cr}(0)=\x_a$}.
\end{definition}
Recall from Appendix \ref{sec. variation} that a critical path for
a variational problem with a boundary must satisfy the Euler
equations and supplemental `transversality' conditions.

\begin{definition}
Let $\Omega:=\Za\times\Ta$. Characterize an integrator on $\Omega$
by $\DO{\Omega}:=\DQ{z}\,\DO{\tr}$ where $\DQ{z}$ is a Gaussian
integrator characterized by (\ref{gaussian}) and $\DO{\tr}$ is the
gamma integrator characterized by (\ref{d-tau}).
\end{definition}

\begin{theorem}\label{main}
Let $\M$ be a real(complex) $m$-dimensional ($m\geq 2$)
paracompact differentiable manifold with a linear connection, and
let $\mathbb{U}$ be a bounded orientable open region in $\M$ with
boundary $\partial{\mathbb{U}}$.\footnote{ The issue of regular
and irregular points of $\partial \mathbb{U}$ will be ignored.
However, I assume sufficient regularity of $\partial \mathbb{U}$
when required.} Let $\Bold{f}$ and $\Bold\varphi$ be elements of
the space of sections or section distributions of the
$(r,s)$-tensor bundle over $\M$.\footnote{All tensors and tensor
distributions are assumed to be well defined, i.e., they are
continuous, differentiable, analytic, bounded, integrable, etc. as
required for any given case.} Assume given the functional
$\mathcal{S}(x(\tau_{a'},z))$ whose associated bilinear form $Q$
satisfies $\mathrm{Re}(Q(x(\tau_{a'},z))>0$ for
$(x(\tau_{a'},z))\neq 0$.

If
$\Bold{\chi}(\x_a\cdot\mathit{\Sigma}(\dual,z))\in\mathcal{F}_R(\Omega)$
where\footnote{From a physics perspective, it may be useful to
express $\Bold{\chi}$ as a functional of $\tau$ and add a term to
$\mathcal{S}(x(\tr,z))$ that restricts $\tau$ to the value
$\tau_{\mathrm{x}_a}^\bot$. However, for our purposes, this is an
unnecessary step.}
\begin{eqnarray}\label{chi}
  \Bold{\chi}(\x_a\cdot\mathit{\Sigma}(\dual,z))
  &:=&\int_{C_+}\theta(\dual-\tau_{a'})
  \Bold{f}(\x_a\cdot\mathit{\Sigma}(\tau_{a'},z))
  \exp\left\{-\mathcal{S}(x(\tau_{a'},z))\right\}\,d\tau_{a'}\notag\\
  \notag\\
  &+&\Bold{\varphi}(\x_a\cdot\mathit{\Sigma}(\dual,z))
  \exp\left\{-\mathcal{S}(x(\dual,z))\right\}\;;
\end{eqnarray}
then, for $\x_a=x(\tau_a)\in \mathbb{U}$,
\emph{\begin{equation}\label{inhomogeneous}
  \Bold{\Psi}(\x_a)
  =\int_\Omega
  \Bold{\chi}(\x_a\cdot\mathit{\Sigma}(\taubot,z))\,\DO{\Omega}
  \end{equation}}
 is a solution of the inhomogeneous PDE
\emph{\begin{equation}\label{inhomogeneous PDE}
   \left.\left[\frac{G^{\alpha\beta}}{4\pi}
    \mathcal{L}_{{\Xa}}\mathcal{L}_{{\Xb}}+\mathcal{L}_{\mathbf{Y}}
    +V(\x)\right]\right|_{\x=\x_a}\Bold{\Psi}(\x_a)=-\Bold{f}(\x_a)
\end{equation}}
with boundary condition \emph{\begin{equation}\label{psi limit}
   \Bold{\Psi}(\x_B)
   =\Bold{\varphi}(\x_B)\;.
\end{equation}}
\end{theorem}

\ \

 The symbol
$\mathcal{L}_{\Xa}$ represents the Lie derivative in the $\Xa$
direction. (Note that it is possible to have highly non-trivial
functions of $\x$ pre-multiplying partial derivative terms due to
the presence of Lie derivatives.) My convention for the step
function of a complex variable is $\theta(|\ti|-|\ti_0|)=0$ if
$|\ti|-|\ti_0|\leq 0$ and $\theta(|\ti|-|\ti_0|)=1$ for
$|\ti|-|\ti_0|>0$. A specific choice of integration path
$C_+\subset\C_+$ is dictated by a particular application, which
usually entails a restriction on $\tr\in\Ta$. For example, in
typical applications either $\tr=\tr^*$ or $\tr=-\tr^*$ so that
$C_+\subseteq\R_+$ or $C_+\subseteq i\R$ respectively.

\vspace{.4in}

\emph{Proof of Theorem \ref{main}.} From the definition of
$\Bold{\Psi}(\x_a)$ and
$\Bold{\chi}(\x_a\cdot\mathit{\Sigma}(\taubot,z))$ and Proposition
\ref{Fubini},
\begin{eqnarray}\label{LPhi}
  L\Bold{\Psi}(\x_a)
  &:=&\left.\left[\frac{G^{\alpha\beta}}{4\pi}\mathcal{L}_{\Xa}\mathcal{L}_{\Xb}
  +\mathcal{L}_{\mathbf{Y}}+V(\x)\right]\right|_{\x=\x_a}\Bold{\Psi}(\x_a)\notag\\
  &=&L\int_{\Ta}(U_{\taubot}\Bold{\varphi})(\x_a)\;\DO{\tr}\notag\\
  &&+L\int_{\Ta}\taurange\theta(\taubot-\tau_{a'})
  (U_{\tau_{a'}}\Bold{f})(\x_a)\;d\tau_{a'}\,\DO{\tr}
\end{eqnarray}
where
\begin{equation}
  (U_{\tau_{a'}}\Bold{f})(\x_a):=
  \int_{\Za}\Bold{f}(\x_a\cdot\mathit{\Sigma}(\tau_{a'},z))
  \exp\left\{-\mathcal{S}(x(\tau_{a'},z))\right\}\,\DQ{z}\;.
\end{equation}

By Proposition \ref{prop. change of variable}, the integrals over
$\Ta$ can be reduced to one dimensional integrals over $C_+$. The
specific form is given in (\ref{reduced tau}). Then assuming the
necessary conditions on the integrands, $L$ can be taken inside
the integrals.

Furthermore,
\begin{lemma}\label{LU} $L$ commutes with $U_{\dual}$.
\end{lemma}

\emph{Proof.} Consider
\begin{equation}
(U_{\dual}\Bold{\varphi})(\x_a)
=\int_{\Za}\Bold{\varphi}(x(\dual,z)
\exp\left\{-\mathcal{S}(x(\dual,z))\right\}\,\DQ{z}\;.
\end{equation}
Taylor expand $\Bold{\varphi}(x(\dual,z)$ and
$\exp\left\{-\mathcal{S}(x(\dual,z))\right\}$ (assuming the
necessary differentiability) about $\x_a$, and use Proposition
\ref{prop. linearity} to interchange the sum and the integral.
Take the terms that depend explicitly on $\x_a$ outside the
integral over $\Za$, apply $L$, and reverse the steps. (Recall
that the resulting functionals have been assumed integrable in
Section \ref{sec. path integrals}.)$\QED$

\begin{lemma}\label{f PDE}
\begin{equation}
  \taurange\theta(\dual-\tau_{a'})
  L(U_{\tau_{a'}}\Bold{f})(\x_a)\;d\tau_{a'}=-\Bold{f}(\x_a)
  -(U_{\dual}\Bold{f})(\x_a)\;.
\end{equation}
\end{lemma}

\emph{Proof.} Scale the time variable by $\dual^{-1}$ in the
parametrization (\ref{tau parametrization}). The space of paths
$\Za$ gets mapped to the space of paths $\widetilde{Z}_a^d$ with
elements $\widetilde{z}(\tr(\ti)/\dual)=\dual^{-1/2}z(\tr(\ti))$,
and the parametrization becomes
\begin{equation}\label{scaled parametrization}
  \left\{ \begin{array}{ll}
    \dot{\Bold{x}}(\tr(\ti)/\dual)
    -\dual\mathbf{Y}(x(\tr(\ti)/\dual))=\\
   \hspace{1in}\dual^{1/2}{\Xa}
   (x(\tr(\ti)/\dual))\dot{\widetilde{\Bold{z}}}
   \,^\alpha(\tr(\ti)/\dual)\\ \\
    x(\tau_a/\dual)=\x_a\;.
 \end{array}
  \right.
\end{equation}
Expand $\Bold{f}(x(\dual,\widetilde{z}))$ about the point $\x_a$
making use of (\ref{scaled parametrization}):
\begin{equation}\label{expansion}
\begin{array}{lll}
  \Bold{f}(\x_a\cdot\mathit{\Sigma}(\dual,\widetilde{z}))
  &=&\Bold{f}(x_a)+\dual^{1/2}
  \mathcal{L}_{\Xa}\Bold{f}(\x_a)\widetilde{z}\,^{\alpha}(1)\\
  &&+\frac{1}{2}\,\dual
  \mathcal{L}_{\Xa}\mathcal{L}_{\Xb}\Bold{f}(\x_a)
  \widetilde{z}\,^{\alpha}(1)\widetilde{z}\,^{\beta}(1)\\
  &&+\dual\mathcal{L}_{\mathbf{Y}}\Bold{f}(\x_a)
  +O(\dual^{3/2})\;.
  \end{array}
\end{equation}
Likewise, the functional $\mathcal{S}(x(\dual,\widetilde{z}))$
becomes
\begin{equation}\label{scaled action}
  \mathcal{S}(x(\dual,\widetilde{z}))
  =\pi Q(x(1,\widetilde{z}))
  -\dual\int_0^1V(x(\ti,\widetilde{z}))\,d\ti
\end{equation}
and
\begin{equation}\label{expanded potential}
\begin{array}{lll}
  V(\x_a\cdot\mathit{\Sigma}(\dual,\widetilde{z}))&=&V(\x_a)
  +\dual^{1/2}\mathcal{L}_{\Xa}V(\x_a)\widetilde{z}\,^{\alpha}(1)
  \\
   &&+\frac{1}{2}\,\dual\mathcal{L}_{\Xa}\mathcal{L}_{\Xb}V(\x_a)
  \widetilde{z}\,^{\alpha}(1)\widetilde{z}\,^{\beta}(1)\\
  &&+\dual\mathcal{L}_{\mathbf{Y}}V(\x_a)
  +O(\dual^{3/2})\;.
  \end{array}
\end{equation}

From Appendix \ref{app. gaussian integrators}---which records the
normalization, mean, and covariance of a Gaussian integrator---and
(\ref{expansion})--(\ref{expanded potential}),
\begin{eqnarray}
(U_{\dual}\Bold{f})(\x_a)&=&\int_{\widetilde{Z}_a^d}
\left\{\Bold{f}(\x_a)+\dual^{1/2}
  \mathcal{L}_{\Xa}\Bold{f}(\x_a)\widetilde{z}\,^{\alpha}(1)\right.
  \notag\\
  &&\hspace{.3in}+\frac{1}{2}\,\dual
  \mathcal{L}_{\Xa}\mathcal{L}_{\Xb}\Bold{f}(\x_a)
  \widetilde{z}\,^{\alpha}(1)\widetilde{z}\,^{\beta}(1)\notag\\
  &&\hspace{.3in}+\left.\dual\mathcal{L}_{\mathbf{Y}}\Bold{f}(\x_a)
  +O(\dual^{3/2})\right\}\notag\\
  &&\hspace{.2in}\times\left\{1+\dual V(\x_a)+O(\dual^2)\right\}
  \,\mathcal{D}\omega(\widetilde{z})\notag\\
  &=&\Bold{f}(\x_a)+\dual\{Lf\}(\x_a)+O(\dual^{3/2})\;.
\end{eqnarray}
The last equality is a consequence of Proposition \ref{linearity}.
It follows from Lemma \ref{prop. semi-group} that
$(U_{\dual}\Bold{f})(\x_a)$ satisfies the PDE
\begin{equation}\label{f}
  \frac{\partial(U_{\dual}\Bold{f})(\x_a)}
  {\partial\tau_{a'}}=L(U_{\dual}\Bold{f})(\x_a)
\end{equation}
with initial condition
\begin{equation}\label{initial f}
  (U_{\dual=0}\Bold{f})(\x_a)=\Bold{f}(\x_a)\;.
\end{equation}
The statement of the lemma is the integral form of the PDE. $\QED$

Using Lemmas \ref{LU} and \ref{f PDE}, equation (\ref{LPhi}) can
be written
\begin{eqnarray}\label{intermediate}
  L\Bold{\Psi}(\x_a)
  =-\Bold{f}(\x_a)-\mathcal{N}\int_{C_+}
  (U_{\ti}\left\{\Bold{f}-L\Bold{\varphi}\right\})(\x_a)
  \,d(\ln\ti)\;.
\end{eqnarray}

\begin{lemma}\label{vanishing}
\begin{equation}
  0=\int_{C_+}
  (U_{\ti}\left\{\Bold{f}-L\Bold{\varphi}\right\})(\x_a)
  \,d(\ln\ti)\;.
\end{equation}
\end{lemma}

\emph{Proof.} By Proposition \ref{Fubini},
\begin{eqnarray}
  \Bold{\Psi}(\x_a)&=&\int_{\Omega}
  \Bold{\chi}(\x_a\cdot\mathit{\Sigma}(\taubot,z))\,\DO{\Omega}
  \notag\\
  &=&\int_{\Ta}\int_{\Za}
  \Bold{\chi}(\x_a\cdot\mathit{\Sigma}(\taubot,z))\,\DQ{z}\,\DO{\tr}
  \notag\\
  &=:&\int_{\Ta}(\overline{U}_{\taubot}\Bold{\chi})(\x_a)\,\DO{\tr}\;.
\end{eqnarray}
From Proposition \ref{scale invariance}, $\DO{\tr}$ is scale
invariant and it therefore verifies the integration by parts
formula (\ref{integration by parts}) for
$\overline{U}_{\taubot}\Bold{\chi}\in\mathcal{F}_R(\Ta)$;
\begin{equation}\label{delta U chi}
  0=\int_{\Ta}\frac{\delta(\overline{U}_{\taubot}\Bold{\chi})(\x_a)}
  {\delta{\tr(\ti)}}\;\DO{\tr}\;.
\end{equation}
But,
\begin{eqnarray}
  \frac{\delta(\overline{U}_{\dual}\Bold{\chi})(\x_a)}
  {\delta{\tr(\ti)}}\notag\\
  &&\hspace{-.7in}=\int_{\Za}\left\{\taurange
  \delta(\dual-\tau_{a'})
  \Bold{f}(\x_a\cdot\mathit{\Sigma}(\tau_{a'},z))
  \exp^{\{-\mathcal{S}(x(\tau_{a'},z))\}}\right.
  \,d\tau_{a'}\nonumber\\
  &&\hspace{-.1in}
  +\left[\Bold{\varphi}'(\x_a\cdot\mathit{\Sigma}(\dual,z))
  -\Bold{\varphi}(\x_a\cdot\mathit{\Sigma}(\dual,z))
  \mathcal{S}'(x(\dual,z))\right]\nonumber\\
  &&\hspace{2.5in}\times\left.\exp^{\{-\mathcal{S}(x(\dual,z))\}}\right\}
  \,\DQ{z}\nonumber\\
  &&\hspace{-.7in}=\int_{\Za}\left\{
  -\Bold{f}(\x_a\cdot\mathit{\Sigma}(\dual,z))\right.\nonumber\\
  &&\hspace{-.2in}
  +\left.\left[\Bold{\varphi}'(\x_a\cdot\mathit{\Sigma}(\dual,z))
  -\Bold{\varphi}(\x_a\cdot\mathit{\Sigma}(\dual,z))
  \mathcal{S}'(x(\dual,z))\right]\right\}\nonumber\\
  &&\hspace{2.5in}\times\exp^{\{-\mathcal{S}(x(\dual,z))\}}
  \,\DQ{z}\nonumber\\
  &&\hspace{-.7in}=:\int_{\Za}\left\{
  -\Bold{f}(\x_a\cdot\mathit{\Sigma}(\dual,z))
  +\Bold{\xi}(\x_a\cdot\mathit{\Sigma}(\dual,z))\right\}
  \exp^{\{-\mathcal{S}(x(\dual,z))\}}
  \,\DQ{z}\nonumber\\
  \notag\\
  &&\hspace{-.7in}=-(U_{\dual}\{\Bold{f}-\Bold{\xi}\})(\x_a)\;.
\end{eqnarray}

From Proposition \ref{prop. semi-group},
\begin{equation}
  U_{\dual}\{\Bold{f}-\Bold{\xi}\}
  =U_{\dual+\langle\delta_{\ti_a},\tr\rangle}\{\Bold{f}-\Bold{\xi}\}
  =U_{\dual}
  (U_{\langle\delta_{\ti_a},\tr\rangle}\{\Bold{f}-\Bold{\xi}\})
\end{equation}
since $\langle\delta_{\ti_a},\tr\rangle=0$. Use the expansion
technique from Lemma \ref{f PDE} to expand $\Bold{\varphi}'$ and
$\mathcal{S}'$ in $\Bold{\xi}$, do the integration over
$\widetilde{Z}^d_a$ as before, and evaluate at
$\langle\delta_{\ti_a},\tr\rangle=0$ to get
\begin{eqnarray}\label{F expansion}
  U_{0}\{\Bold{f}-\Bold{\xi}\}(\x_a)
  &=&\Bold{f}(\x_a)-\left.\left[
  \mathcal{L}_{\mathbf{Y}}
  +\frac{G^{\alpha\beta}}{4\pi}\mathcal{L}_{\Xa}\mathcal{L}_{\Xb}
  +V(\x)\right]\right|_{\x=\x_a}\Bold{\varphi}(\x_a)\nonumber\\
  &=&\Bold{f}(\x_a)-L\Bold{\varphi}(\x_a)\;.
\end{eqnarray}

Hence, (\ref{delta U chi}) can be rewritten as
\begin{equation}\label{rewritten proof}
  0=\int_{\Ta}(U_{\taubot}\{\Bold{f}-L\Bold{\varphi}\})(\x_a)\;\DO{\tr}\;.
\end{equation}
Since this reduces to a one dimensional integral according to
(\ref{reduced tau}), the lemma is verified.\footnote{Intuitively,
this makes sense---at least in the case where
$(U_{\ti}\{\Bold{f}-L\Bold{\varphi}\})(\x_a)$ solves the diffusion
equation, i.e. $\ti\in\R_+$. The integrator is scale invariant,
and $(U_{\ti}\{\Bold{f}-L\Bold{\varphi}\})(\x_a)\rightarrow 0$ as
$\ti\rightarrow\infty$.} $\QED$

So, by Lemma \ref{vanishing}, equation (\ref{intermediate})
reduces to the inhomogeneous PDE
\begin{equation}
  L\Bold{\Psi}(\x_a)=-\Bold{f}(\x_a)\;.
\end{equation}

To verify the boundary condition, note that the definition of
$\taubot$ implies that $\x_a\rightarrow \x_B\in\partial
\mathbb{U}$ implies $\taubot\rightarrow 0$.\footnote{This is
easily seen, for example, by using normal coordinates near the
boundary and recalling that $x_{cr}$ is a critical path. As soon
as the starting point lies on the boundary, $\tau_{\x_B}^\bot=0$
for any endpoint on the boundary.} The boundary condition follows
readily:
\begin{eqnarray}
  \Bold{\Psi}(\x_a)|_{\x_a=\x_B}
  &=&\int_{\widetilde{\Omega}}\taurange
  \Bold{\chi}(\x_B\cdot\mathit{\Sigma}(\tau_{a'},\widetilde{z}),0)
  \exp\left\{-\mathcal{S}(x(\tau_{a'},\widetilde{z}))\right\}\;d\tau_{a'}
  \,\mathcal{D}\widetilde{\Omega}\nonumber\\
&=&\Bold{\varphi}(\x_B) \int_{\widetilde{\Omega}}
\exp^{\left\{0\int_0^{1}V(x(\ti,\widetilde{z}))\,d\ti\right\}}
  \exp^{\left\{-\pi Q(x(1,\widetilde{z}))\right\}}
  \;\mathcal{D}\widetilde{\Omega}\nonumber\\
  &=&\Bold{\varphi}(\x_B)\;.
\end{eqnarray}
The second equality holds because
$\x_B\cdot\mathit{\Sigma}(0,\widetilde{z})=\x_B$ is independent of
$\widetilde{z}$, and the third equality follows from the
normalization of the integrators.

This completes the proof of the theorem. $\QED$

\emph{Remarks about Theorem \ref{main}}:
\begin{description}

\item[--] Note that there may be more than
one critical path and, hence, more than one
$\tau_{\mathrm{x}_a}^\bot$. In that case, (\ref{inhomogeneous})
should then include a sum over all $\tau_{\mathrm{x}_a}^\bot$.

\item[--] If the boundary of $\mathbb{U}$ is at infinity and
$\Bold{\varphi}(\x_B)=0$, then (\ref{inhomogeneous}) can be
written as the Fourier/Laplace transform of a path integral which
solves an inhomogeneous parabolic equation
  associated with (\ref{inhomogeneous PDE}).\footnote{If the potential
  $V(\x)$ has no non-vanishing constant term, then the
  associated parabolic PDE has a different potential, viz.
  $V'(\x)=V(\x)-2\pi\imath \mathcal{E}$.}
  This is a well known result for the case of elliptic PDEs.

  However, when $\M$
  corresponds to the phase space of a dynamical system, the
  path integral is more naturally rewritten in terms of a
  Lagrange multiplier by employing the map $R:\Ta\rightarrow
  \mathit{\Lambda}$ which maps $\tr\mapsto\lambda$ by
\begin{equation}
  \tr(\ti)=\int_{\ti_a}^{\ti_b}\theta(\ti-\ti')
  \lambda(\ti')\,d\ti'\;.
\end{equation}Consequently, the Fourier transform interpretation is
  no longer evident.

   Under this map, the action then takes the form of a constrained system and the
   domain of integration reverts back to $[\ti_a,\ti_b]$. This
   justifies expressing (\ref{inhomogeneous}) in terms of a path
   integral over $\Ta$ instead of an equivalent but simpler integral over $C_+$.
In this sense, the phase space representation
  of (\ref{inhomogeneous}) appears to be more general and, therefore,
  perhaps more useful.

\item[--] For $\Bold{f}=1$, $\Bold{\varphi}=0$, and suitable
$V(\x)$,
\begin{equation*}
  \Bold{\Psi}(\x_a)=\int_{\Ta}\taubot\DO{\tr}
  =:{\langle\taubot\rangle}_{\Ta}
\end{equation*}
so that $L{\langle\taubot\rangle}_{\Ta}=-1$ with boundary
condition ${\langle \tau^{\bot}_{\x_B}\rangle}_{\Ta}=0$. (More
precisely, this follows after integrating (\ref{interior G}) over
$x_{a'}$.)

\item[--] The (pointed) path integrals can be transformed into (pointed) loop
integrals by replacing the parameter interval
$\mathbb{T}=[\ti_a,\ti_b]$ with $S^1$ and requiring $\tr$ to be a
homeomorphism.

\item[--] The case of simply connected, compact $\mathbb{U}$ without boundary can be
included in the theorem by taking $\M$ to be a simply connected,
compact manifold without boundary and then deforming the boundary
$\partial \mathbb{U}$ to a point (without intersecting $\x_a$, the
point of interest). This amounts to putting $\Bold{\varphi}=0$,
and letting $\taubot$ represent the period of a critical loop in
$\mathbb{U}$.

\item[--] It is important to note that the proof of the theorem doesn't require
  that $\partial{\mathbb{U}}$ be closed: Nor does it address the boundary type and
  boundary conditions necessary for existence and uniqueness of
  the solutions. For scalar functions on $\R^n$,
  typical parabolic and hyperbolic PDEs require an open
  boundary with Cauchy
  initial condition---possibly supplemented with
  Dirichlet/Neumann boundary conditions.
  On the other hand, elliptic PDEs
  require a closed boundary with Dirichlet/Neumann boundary conditions.
  However, the boundary type
  and boundary conditions necessary for existence and uniqueness of
  solutions for the generalized PDEs considered here are open questions.

\end{description}

\subsection{Kernels and Eigenfunctions}
In practice, finding the functional $\Bold{\chi}$ for general
boundary conditions can be difficult. The task can be simplified
by constructing kernels of (\ref{inhomogeneous PDE}), denoted by
$\K$, appropriate for Dirichlet/Neumann boundary conditions.

\subsubsection{Dirichlet Kernels}\label{Dirichlet kernels}

\begin{lemma}
The Dirichlet elementary kernel is given by
\begin{eqnarray}\label{interior G}
  \K_\mathbb{U}^{(D)}(\x_a,\x_{a'})&:=&\int_\Omega\taurange
  \theta(\taubot-\tau_{a'})
  \Bold{\delta}(x(\tau_{a'},z),\x_{a'})\nonumber\\
  &&\hspace{.4in}\times
  \exp{\left\{-\mathcal{S}(x(\tau_{a'},z))\right\}}\;d\tau_{a'}\,\DO{\Omega}
\end{eqnarray}
where $\x_a,\x_{a'}\in\U$ and $\Bold{\delta}(\x_a,\x_{a'})$
denotes a Dirac bitensor composed of Kronecker delta symbols
(which are collectively denoted by $\mathbf{1}$ since the specific
form is not necessary for our purposes) and the scalar Dirac delta
function $\delta(\x_a,\x_{a'})$.
\end{lemma}

\emph{Proof.} It follows immediately from the theorem that
$\K_\mathbb{U}^{(D)}$ is the elementary solution of the
inhomogeneous PDE. Moreover, $\K_\mathbb{U}^{(D)}(\x_B,\x_{a'})=0$
since $\x_a\rightarrow \x_B$ implies $\taubot\rightarrow 0$.
Evidently, $\K_\mathbb{U}^{(D)}(\x_a,\x_{a'})$ is the elementary
kernel of the PDE which vanishes on the boundary.$\QED$

\begin{lemma}
The Dirichlet boundary kernel is given by
\begin{eqnarray}\label{boundary G}
\K_{\partial}^{(D)}(\x_a,\x_B)&:=&\int_\Omega
\Bold{\delta}(x(\taubot,z),\x_B)
\exp{\left\{-\mathcal{S}(x(\taubot,z))\right\}} \;\DO{\Omega}
\end{eqnarray}
where $\x_B\in\partial \mathbb{U}$.
\end{lemma}

\emph{Proof.} According to the theorem,
$\K_{\partial}^{(D)}(\x_a,\x_B)$ is a kernel of the homogeneous
PDE and $\K_{\p}^{(D)}(\x_{B'},\x_B)=\Bold{\delta}(\x_{B'},\x_B)$
for $\x_{B'}\in\p\mathbb{U}$.$\QED$

\begin{corollary}
For Dirichlet boundary conditions $\Bold{\varphi}(\x_B)$, the
solution of the inhomogeneous PDE can be written
\begin{equation}\label{kernel solution}
  \Bold{\Psi}(\x_a)
  =\int_{\mathbb{U}}\K_\mathbb{U}^{(D)}(\x_a,\x_{a'})
  \Bold{f}(\x_{a'})\;d\x_{a'}
  +\int_{\partial{\mathbb{U}}}\K_{\partial}^{(D)}(\x_a,\x_B)
  \Bold{\varphi}(\x_B)\;d\x_B\;
\end{equation}
where $d\x_a$ is the volume element on $\mathbb{U}$, and $d\x_B$
is the associated Leray form on $\partial \mathbb{U}$.
\end{corollary}

\emph{Proof.} The corollary follows from the preceding lemmas.
$\QED$

Notice that the boundary kernel is a path integral over paths
starting at $\x_a$ at $\tau_a$ and ending at \emph{some} point on
the boundary at time $\taubot$. It can be expressed indirectly in
terms of paths starting at $\x_a$ at $\tau_a$ and ending at the
specific point $x_{cr}(\taubot)=:\x_a^\bot\in\p\mathbb{U}$. This
follows because
$\K_{\p}^{(D)}(\x_{B'},\x_B)=\Bold{\delta}(\x_{B'},\x_B)$ for
$\x_{B'}\in\p\mathbb{U}$. (The point is it doesn't cost anything
to propagate along the boundary.) Hence,
\begin{eqnarray}\label{crtical kernel}
  \K_{\partial}^{(D)}(\x_a,\x_a^\bot)&=&\int_{\p\mathbb{U}}
  \K_{\partial}^{(D)}(\x_a,\x_B)\K_{\partial}^{(D)}(\x_B,\x_a^\bot)
 \,d\x_B\nonumber\\
  &=&\int_{\p\mathbb{U}}
  \K_{\partial}^{(D)}(\x_a,\x_B)\Bold{\delta}(\x_B,\x_a^\bot)
  \,d\x_B\nonumber\\
  &=&\int_\Omega
\Bold{\delta}(x(\taubot,z),\x_a^\bot)
\exp{\left\{-\mathcal{S}(x(\taubot,z))\right\}}
\;\DO{\Omega}\;.\nonumber\\
\end{eqnarray}
This form is particularly useful in calculations when one expands
about the critical path(s).

For later use, it is convenient to rewrite (\ref{interior G}) for
$\mathbb{U}$ embedded in a non-compact $\M$ for the special cases
when $\tr=\tr^*$ or $\tr=-\tr^*$.

\begin{proposition}\label{boundary trick}
\begin{eqnarray}
\K_\mathbb{U}^{(D)}(\x_a,\x_{a'})&=&\int_\Omega
\left[\int_{\widehat{\R}} \Bold{\delta}(x(\tau_{a'},z),\x_{a'})
\exp{\left\{-\mathcal{S}(x(\tau_{a'},z))\right\}}\right.d\tau_{a'}
\,\DO{\Omega}\notag\\
&&-\int_\Omega \int_{\widehat{\R}}
\Bold{\delta}(\widetilde{x}(\tau_{a'},z),\widetilde{\x}_{a'})
\exp{\left\{-\mathcal{S}(\widetilde{x}(\tau_{a'},z);s) \right\}}
d\tau_{a'}\;\DO{\Omega}\notag\\
\end{eqnarray}
where $\widetilde{x}(\tau_{a'},\cdot)
=\sigma(\x_a)\cdot\mathit{\Sigma}(\tau_{a'},\,\cdot\,)$ ,
$\sigma(\x_a)=x_{cr}(\taubot)$, and $\widehat{\R}$ denotes $\R_+$
or $i\R$ depending on whether $\tr=\tr^*$ or $\tr=-\tr^*$.
\end{proposition}

\emph{Proof.} For $\tr=\tr^*$,
\begin{eqnarray}\label{non-compact K}
\K_\mathbb{U}^{(D)}(\x_a,\x_{a'})&=&\int_\Omega
\left[\int_0^{\infty} \Bold{\delta}(x(\tau_{a'},z),\x_{a'})
\exp{\left\{-\mathcal{S}(x(\tau_{a'},z))\right\}}\right.d\tau_{a'}
\nonumber\\
  &&\hspace{.2in}-\left.\int_{\taubot}^{\infty}
\Bold{\delta}(x(\tau_{a'},z),\x_{a'})
\exp{\left\{-\mathcal{S}(x(\tau_{a'},z))\right\}}d\tau_{a'}\right]
\;\DO{\Omega}\nonumber\\
&=:&\K_{\infty}^+(\x_a,\x_{a'})-\mathbf{F}_\mathbb{U}^+(\x_a,\x_{a'})\;;
\end{eqnarray}
and for $\tr=-\tr^*$,
\begin{eqnarray}
\K_\mathbb{U}^{(D)}(\x_a,\x_{a'})
&=&[\K_{\infty}^+(\x_a,\x_{a'})-\mathbf{F}_\mathbb{U}^+(\x_a,\x_{a'})]
+[\K_{\infty}^-(\x_a,\x_{a'})-\mathbf{F}_\mathbb{U}^-(\x_a,\x_{a'})]\;,
\nonumber\\
\end{eqnarray}
where
\begin{eqnarray}
\lefteqn{\K_{\infty}^-(\x_a,\x_{a'})-\mathbf{F}_\mathbb{U}^-(\x_a,\x_{a'})}
\nonumber\\
&:=&\int_\Omega \left[\int_{-\infty}^{0}
\Bold{\delta}(x(\tau_{a'},z),\x_{a'})
\exp{\left\{-\mathcal{S}(x(\tau_{a'},z))\right\}}\right.d\tau_{a'}
\nonumber\\
  &&\hspace{.2in}-\left.\int^{-\taubot}_{-\infty}
\Bold{\delta}(x(\tau_{a'},z),\x_{a'})
\exp{\left\{-\mathcal{S}(x(\tau_{a'},z))\right\}}d\tau_{a'}\right]
\;\DO{\Omega}\;.\nonumber\\
\end{eqnarray}
Let $\K_{\infty}$ denote $\K_{\infty}^+$ or
$[\K_{\infty}^++\K_{\infty}^-]$ and $\mathbf{F}_\mathbb{U}$ denote
$\mathbf{F}_\mathbb{U}^+$ or
$[\mathbf{F}_\mathbb{U}^++\mathbf{F}_\mathbb{U}^-]$ depending on
whether $\mathrm{Im}\tr=0$ or $\mathrm{Re}\tr=0$ respectively.
According to the theorem, $\K_{\infty}(\x_a,\x_{a'})$ is the
elementary kernel for the boundary at infinity since
$\x_{a}\rightarrow \infty$ implies $|\taubot|\rightarrow
\infty$.\footnote{More precisely, since the paths are $L^{2,1}$,
their energy, and hence velocity, are bounded. This requires
$|\tau_{\mathrm{x}_a}^\bot|\rightarrow\infty$ as
$\x_a\rightarrow\infty$.} It follows that  $\mathbf{F}_\mathbb{U}$
must be the kernel to the homogeneous PDE with boundary condition
\begin{equation*}
 \mathbf{F}_\mathbb{U}(\x_B,\x_{a'})=\K_{\infty}(\x_B,\x_{a'})\;.
\end{equation*} (Roughly speaking,
$\mathbf{F}_\mathbb{U}$ satisfies the homogeneous PDE because
$x_{cr}(\tau_{a'})\notin \mathbb{U}$ for
$\tau_{a'}>\tau_{\mathrm{x}_a}^\bot$.)

$\mathbf{F}_\mathbb{U}$ can be rewritten by introducing a point
transformation $\sigma:\mathbb{U}\rightarrow \mathbb{U}$ by
$\x_a\mapsto \sigma(\x_a)=x_{cr}(\taubot)=\x_a^\bot$ (which is a
point on $\p\mathbb{U}$). Consequently, $\taubot\mapsto
\tau_{\sigma(\x_a)}^\bot=0$ under this transformation.

In order to leave the boundary conditions unchanged for paths
whose end-points are restricted by the delta functional in
$\mathbf{F}_\mathbb{U}$, it is necessary to simultaneously
transform $\x_{a'}\rightarrow\widetilde{\x}_{a'}$ such that
$\widetilde{\x}_{a'}=\sigma^{-1}(\x_{a'})$. To see this, let
$\widetilde{\mathit{\Sigma}}$ denote the transformation from
$\x_a$ to the final point $\widetilde{\x}_{a'}$ and
write\begin{equation} \widetilde{x}(\tau_{a'},\cdot)
=\sigma(\x_a)\cdot\mathit{\Sigma}(\tau_{a'},\,\cdot\,)
=\x_a\cdot\widetilde{\mathit{\Sigma}}(\tau_{a'},\,\cdot\,)
=\widetilde{\x}_{a'}
\end{equation}
so that $\widetilde{\mathit{\Sigma}}=\mathit{\Sigma}\circ\sigma$.
Then, for paths with transformed initial point $\sigma(\x_a)$ and
fixed end-point $\widetilde{\x}_{a'}$,
\begin{equation}
\sigma(\widetilde{x}(\tau_{a'},\,\cdot\,))
=\sigma(\widetilde{\x}_{a'}) =\x_{a'} =x(\tau_{a'},\,\cdot\,)
=\x_a\cdot\mathit{\Sigma}(\tau_{a'},\,\cdot\,)\;.
\end{equation}
In particular,
\begin{equation}
  \sigma(\widetilde{x}_{cr}(\taubot,\,\cdot\,))
=\x_a^\bot
\end{equation}
remains consistent with the the parametrization and boundary
conditions.

For paths that contribute to $\mathbf{F}_\mathbb{U}$, it follows
that
 \begin{eqnarray}
  \lefteqn{\mathbf{F}_{\mathbb{U}}
  (\sigma(\x_a),\sigma^{-1}(\x_{a'}))}\nonumber\\
  &=&\int_\Omega \int_{\widehat{\R}}
\Bold{\delta}(\widetilde{x}(\tau_{a'},z),\widetilde{\x}_{a'})
\exp{\left\{-\mathcal{S}(\widetilde{x}(\tau_{a'},z);s)
\right\}} d\tau_{a'}\;\DO{\Omega}\nonumber\\
\end{eqnarray}
where $\widehat{\R}$ denotes $\R_+$ or $i\R$ depending on
$\mathbf{F}_{\mathbb{U}}$. Thus
$\mathbf{F}_{\mathbb{U}}(\sigma(\x_a),\sigma^{-1}(\x_{a'}))$
satisfies the homogeneous PDE, and, since $\sigma\rightarrow
\mathit{Id}$ when $\x_a\rightarrow \x_B$,
$\mathbf{F}_{\mathbb{U}}(\sigma(\x_B),\sigma^{-1}(\x_{a'}))
=\K_{\infty}(\x_B,\x_{a'})$. Therefore,
$\mathbf{F}_{\mathbb{U}}(\sigma(\x_a),\sigma^{-1}(\x_{a'}))$ is
equivalent to $\mathbf{F}_{\mathbb{U}}(\x_a,\x_{a'})$ since it
satisfies the same PDE and boundary conditions (assuming
uniqueness).$\QED$

\subsubsection{Neumann Kernels}

The Neumann elementary kernel $\K_\mathbb{U}^{(N)}$ will be
defined for $\mathbb{U}$ embedded in non-compact $\M$.
\begin{lemma}
Making use of the decomposition introduced in (\ref{non-compact
K}), the Neumann elementary kernel is given by
\begin{eqnarray}\label{Neumann kernel}
  \K_{\mathbb{U}}^{(N)}(\x_a,\x_{a'})
  &:=&\K_{\infty}(\x_a,\x_{a'})
  +\mathbf{F}_{\mathbb{U}}(\x_a,\x_{a'})\;.
\end{eqnarray}
\end{lemma}

\emph{Proof.} Clearly $\K_{\mathbb{U}}^{(N)}$ satisfies the
inhomogeneous PDE. For the boundary condition,
\begin{eqnarray}
   \left.\nabla_{\mathbf{n}_{\partial}}\K_{\mathbb{U}}^{(N)}(\x_a,\x_{a'})
   \right|_{\partial \mathbb{U}}
  &=& \left.\nabla_{\mathbf{n}_{\partial}}\left[\K_{\infty}(\x_a,\x_{a'})
  +\mathbf{F}_{\mathbb{U}}(\x_a,\x_{a'})\right]
  \right|_{\partial \mathbb{U}}\nonumber\\
  &=&\nabla_{\mathbf{n}_{\partial}}\K_{\infty}(\x_B,\x_{a'})
  -\nabla_{\mathbf{n}_{\partial}}\mathbf{F}_{\mathbb{U}}(\x_B,\x_{a'})
  \nonumber\\
  &=&0
\end{eqnarray}
where ${\mathbf{n}_{\partial}}$ is a unit normal to the boundary
in the direction of the interior of $\mathbb{U}$. The second
equality follows because, from the construction,
$\nabla_{\mathbf{n}_{\partial}} x(\tau_{a'}\,,\,\cdot)|_{\partial
\mathbb{U}}$ contributes opposite signs for paths which contribute
to $\K_{\infty}$ and $\mathbf{F}_{\mathbb{U}}$ respectively. The
third equality follows from (\ref{non-compact K}) recalling that
$\taubot\rightarrow 0$ when $\x_a\rightarrow\x_B$.$\QED$

Note that (\ref{Neumann kernel}) can be written
\begin{eqnarray}
  \K_{\mathbb{U}}^{(N)}(\x_a,\x_{a'})
  &=&\K_{\infty}(\x_a,\x_{a'})
  +\mathbf{F}_{\mathbb{U}}(\sigma(\x_a),\sigma^{-1}(\x_{a'}))\;.
\end{eqnarray}

\begin{lemma}
The Neumann boundary kernel is given by
\begin{eqnarray}
  \K_{\partial}^{(N)}(\x_a,\x_B)&:=&\int_\Omega
    \Bold{\theta}(x(\taubot,z),\x_B)
  \exp{\left\{-\mathcal{S}(x(\taubot,z))\right\}}\,\DO{\Omega}
\end{eqnarray}
where $\nabla_{\mathbf{n}_{\partial}}\Bold{\theta}(\x,\x_B)
=\Bold{\delta}(\x,\x_B)$ and
\begin{equation}
\Bold{\theta}(\x,\x_B):=\left\{\begin{array}{ll}
\mathbf{0}\hspace{.5in}\mbox{for}\hspace{.5in}\x\preceq\x_B\\
\mathbf{1}\hspace{.5in}\mbox{for}\hspace{.5in}\x\succ \x_B
\end{array}\right.\;.
\end{equation}
\end{lemma}
The ordering is with respect to a foliation induced by a Gauss
normal coordinate system relative to the boundary.

\emph{Proof.} The theorem ensures that $\K_{\partial}^{(N)}$ is a
kernel of the homogeneous PDE. Also,
$\left.\nabla_{\mathbf{n}_{\partial}}\K_{\partial}^{(N)}
\right|_{\p\mathbb{U}}
=\left.\K_{\partial}^{(D)}\right|_{\p\mathbb{U}}$, because the
$\Bold{\theta}$ term vanishes on the boundary since
$\taubot\rightarrow 0$ as $\x_a\rightarrow \x_B$. So the boundary
condition is
$\nabla_{\mathbf{n}_{\partial}}\K_{\partial}^{(N)}(\x_{B'},\x_B)
=\Bold{\delta}(\x_{B'},\x_B)$.$\QED$

\begin{corollary}
For Neumann boundary conditions
$\nabla_{\mathbf{n}_{\partial}}\Bold{\varphi}(\x_B)$ the solution
of the inhomogeneous PDE can be written (up to a possible
constant)
\begin{equation}
  \Bold{\Psi}(\x_a)
  =\int_{\mathbb{U}}\K_\mathbb{U}^{(N)}(\x_a,\x_{a'})
  \Bold{f}(\x_{a'})\;d\x_{a'}
  +\int_{\partial{\mathbb{U}}}\K_{\partial}^{(N)}(\x_a,\x_B)
  \nabla_{\mathbf{n}_{\partial}}\Bold{\varphi}(\x_B)\;d\x_B\;.
\end{equation}
\end{corollary}

\subsubsection{Eigenfunctions and Eigenvalues}
Notice that it is not possible (using the theorem) to determine
eigenfunctions satisfying $(L-\lambda)\Bold{\Psi}=0$ with
vanishing Dirichlet boundary conditions since this would require
both $\Bold{f}=0$ and $\Bold{\varphi}=0$. However, this tack can
be used for eigenfunctions with non-vanishing Dirichlet boundary
conditions, i.e. $\Bold{\varphi}\neq 0$.

Eigenfunctions that satisfy $(L-\lambda)\Bold{\Psi}=0$ \emph{which
vanish on the boundary} can be represented  in terms of the
elementary kernel with the choice of $\Bold{f}=\lambda\Bold{\Psi}$
and $\Bold{\varphi}=0$. Then (\ref{kernel solution}) becomes a
Fredholm Equation of the second kind;
\begin{equation}
  \Bold{\Psi}(\x_a)
  =\lambda\int_{\mathbb{U}}\K_\mathbb{U}^{(D)}(\x_a,\x_{a'})
  \Bold{\Psi}(\x_{a'})\;d\x_{a'}\;.
\end{equation}
The allowed eigenvalues can be determined by established methods.


\section{Conclusion}
The construction of a path integral that is a solution to a
general class of linear second order PDEs was presented.
Specialization leading to path integral solutions of elliptic,
parabolic, and hyperbolic PDEs with Dirichlet/Neumann boundary
conditions will be developed in a subsequent paper.

This work can be expanded in several directions. One could try to
generalize the path integral to solve higher order or quasilinear
PDEs. It would be potentially profitable to check if this
construction offers any advantages regarding numerical methods. In
a physics context, the integrals that have been constructed are
based on configuration space. These can be recast in terms of
integrals based on phase space yielding a (perhaps) more
fundamental or useful representation. Finally, it is important to
extend this work to Grassmann variables and fields.

More significantly, it remains to determine existence and
uniqueness of the solutions for the general class of PDEs
considered here. This is no small task. It is hoped that the
unifying construction presented here will aid in the effort; and
maybe afford some insight into PDEs in general.

\vspace{.3in} \noindent\textbf{Acknowledgment} I thank C.
DeWitt-Morette for helpful suggestions and discussions.

\appendix
\section{Path Integral Properties}\label{app. properties}
Recall that $\mu$ is a complex Borel measure and $\Theta(x,x')$
and $Z(x')$ are assumed to be continuous, bounded, and
$\mu$-integrable functionals. Many properties of
$\int_{X}\,\mathcal{D}_{\Theta,Z}x$ derived in this appendix are a
consequence of analogous properties of $\int_{X'}\,d\mu$.
\begin{proposition}[Linearity]\label{prop. linearity}
The integral operator $\int_{X}\,\mathcal{D}_{\Theta,Z}x$ is a
linear operator on $\mathcal{F}(X)$.\end{proposition}
\emph{Proof.}
\begin{eqnarray}\label{linearity}
  \int_{X}(aF_{\mu}+bF_{\nu})(x)\,\mathcal{D}_{\Theta,Z}x
 &=&\int_{X'}Z(x')\,(ad\mu+bd\nu)(x')\nonumber\\
 &=&a\int_{X'}Z(x')\,d\mu(x')+b\int_{X'}Z(x')\,d\nu(x')\nonumber\\
 &=&a\int_{X}F_{\mu}(x)\,\mathcal{D}_{\Theta,Z}x
 +b\int_{X}F_{\nu}(x)\,\mathcal{D}_{\Theta,Z}x\;.\nonumber\\
\end{eqnarray}
$\QED$

\begin{proposition}[Order of integration]\label{prop. order interchange}
The order of integration over $X$ and $X'$ can be
interchanged.
\end{proposition}
\emph{Proof.} From Definitions \ref{def1}-\ref{def3}, it follows
readily that
\begin{eqnarray}\label{order interchange}
   \int_{X}\left[\int_{X'}\Theta(x,x')\,d\mu(x')\right]\,\mathcal{D}_{\Theta,Z}x
   &=&\int_{X}F_{\mu}(x)\,\mathcal{D}_{\Theta,Z}x\notag\\
   &=&\int_{X'}\left[\int_{X}\Theta(x,x')\,\mathcal{D}_{\Theta,Z}x\right]\,d\mu(x')\;.
\end{eqnarray}
$\QED$

\begin{proposition}[Fubini]\label{prop. Fubini}
Suppose that $X$ is the disjoint union of two separable Banach
spaces, i.e., $X=X_1\cup X_2$ and $X_1\cap X_2=\emptyset$. The
dual space $X'$ likewise decomposes. If
$\Theta(x,x')=\Theta((x_1,x_2),(x_1',x_2'))
=\Theta_1(x_1,x_1')\cdot\Theta_2(x_2,x_2')$ and
$Z(x')=Z((x_1',x_2'))=Z_1(x_1')\cdot Z_2(x_2')$, then
\begin{eqnarray}\label{Fubini}
  \int_XF_{\mu}(x)\,\mathcal{D}_{\Theta,Z}x
  &=&\int_{X_1}\left[\int_{X_2}F(x_1,x_2)
  \,\mathcal{D}_{\Theta_2,Z_2}x_2\right]\,\mathcal{D}_{\Theta_1,Z_1}x_1\notag\\
  &=&\int_{X_2}\left[\int_{X_1}F(x_1,x_2)
  \,\mathcal{D}_{\Theta_1,Z_1}x_1\right]\,\mathcal{D}_{\Theta_2,Z_2}x_2\;.
\end{eqnarray}\end{proposition}

\emph{Proof.}
\begin{eqnarray}
  \int_XF_{\mu}(x)\,\mathcal{D}_{\Theta,Z}x
  &=&\int_{X'}Z(x')\,d\mu(x')\nonumber\\
  &=&\int_{X_1'}\int_{X_2'}Z_1(x_1')Z_2(x_2')\,d\mu(x_1')\,d\mu(x_2')
  \nonumber\\
  &=&\int_{X_1'}\int_{X_2'}\int_{X_1}\int_{X_2}
  \Theta_1(x_1,x_1')\Theta_2(x_2,x_2')\nonumber\\
  &&\hspace{1in}\times\,\mathcal{D}_{\Theta_1,Z_1}x_1
  \,\mathcal{D}_{\Theta_2,Z_2}x_2\,d\mu(x_1')\,d\mu(x_2')\nonumber\\
&=&\int_{X'}\int_{X_1}\int_{X_2}\Theta(x,x')
  \,\mathcal{D}_{\Theta_2,Z_2}x_2\,\mathcal{D}_{\Theta_1,Z_1}x_1
  \,d\mu(x')\nonumber\\
&=&\int_{X_1}\int_{X_2}\int_{X'}\Theta(x,x')\,d\mu(x')
  \,\mathcal{D}_{\Theta_2,Z_2}x_2\,\mathcal{D}_{\Theta_1,Z_1}x_1\nonumber\\
  &=&\int_{X_1}\left[\int_{X_2}F(x_1,x_2)\,\mathcal{D}_{\Theta_2,Z_2}x_2\right]
  \,\mathcal{D}_{\Theta_1,Z_1}x_1
\end{eqnarray}
The fourth equality is a consequence of the Fubini theorem for
$\mu$. The order of integration over $X'$ and $X$ in the fifth
equality can be interchanged according to Proposition \ref{prop.
order interchange}. $\QED$

Note that $\mathcal{D}_{\Theta_1,Z_1}x_1$ and
$\mathcal{D}_{\Theta_2,Z_2}x_2$ are possibly different integrators
depending on $\Theta_1$, $\Theta_2$ and $Z_1$, $Z_2$.

\begin{corollary}\label{finite interchange}
For $X$ an infinite dimensional separable Banach space and $\M$ a
finite dimensional manifold,
\begin{equation}
  \int_{X}\int_{\M}=\int_{\M}\int_{X}\;.
\end{equation}
\end{corollary}

\emph{Proof.} Definition \ref{def1} can be used to define a
translation invariant integrator for finite dimensional $X=\R^n$
by
\begin{equation}
  \int_{\R^n}\exp^{\{-\pi W_{\R^n}^{-1}(\mathbf{u})-2\pi
i\langle\mathbf{u}',\mathbf{u}\rangle_{\R^n}\}}
|\mathrm{det}\,W_{\R^n}|^{-1/2}\,d\mathbf{u} =\exp^{\{-\pi
W_{\R^n}(\mathbf{u'})\}}
\end{equation}
where $W_{\R^n}$ is a positive definite invertible quadratic form
on $\R^n$. This integrator can be extended to manifolds in the
usual way. Then if $X_2=\R^n$ in Proposition \ref{prop. Fubini},
$\int_{X}\int_{\R^n}=\int_{\R^n}\int_{X}$ and by extension, the
corollary follows.$\QED$

\begin{proposition}[Mean value]\label{prop. mean value1}
Version 1: If there exists an $\overline{x}\in X$ such that
$\Theta(\overline{x},\cdot)=Z(\cdot)$, then
\begin{equation}
  F_{\mu}(\overline{x})=\int_XF_{\mu}(x)\,\mathcal{D}_{\Theta,Z}x\;.
\end{equation}

Version 2: Let $\langle x\rangle :=
\int_Xx\,\mathcal{D}_{\Theta,Z}x$ and define
$\widetilde{\Theta}(\langle x\rangle,\cdot):=Z(\cdot)$. Then,
\begin{equation}
  \widetilde{F_{\mu}}(\langle x\rangle)
  =\int_XF_{\mu}(x)\,\mathcal{D}_{\Theta,Z}x\;.
\end{equation}
\end{proposition}
\emph{Proof.} Version 1:\begin{eqnarray}
  F_{\mu}(\overline{x})&=&\int_{X'}\Theta(\overline{x},x')d\mu(x')
  \nonumber\\
  &=&\int_{X'}Z(x')d\mu(x')\nonumber\\
  &=&\int_XF_{\mu}(x)\,\mathcal{D}_{\Theta,Z}x\;.
\end{eqnarray}
Version 2:
\begin{eqnarray}
  \widetilde{F_{\mu}}(\langle x\rangle)
  &:=&\int_{X'}\widetilde{\Theta}(\langle x\rangle,x')d\mu(x')\nonumber\\
  &=&\int_{X'}Z(x')d\mu(x')\nonumber\\
  &=&\int_XF_{\mu}(x)\,\mathcal{D}_{\Theta,Z}x\;.
\end{eqnarray}
$\QED$

\begin{proposition}[Change of variable]\label{prop. change of variable}
Let $M:X\rightarrow Y$ and $R:Y'\rightarrow X'$ be
diffeomorphisms. Under the change of variable $x\mapsto M(x)=y$,
\begin{equation}\label{change of variable}
  \int_{Y}F_{\mu}(y)\,\mathcal{D}_{\overline{\Theta},\overline{Z}}y
   =\int_{X}F_{\mu}(M(x))\,\mathcal{D}_{\Theta,Z}x
\end{equation}
where $\overline{\Theta}$ and $\overline{Z}$ are defined in
(\ref{theta bar}) and (\ref{Z bar}).

\end{proposition}
\emph{Proof.} Note that
\begin{eqnarray}\label{F circle M}
  (F_{\mu}\circ M)(x)
  &=&\int_{Y'}\overline{\Theta}(M(x),y')\,d\mu(y')\nonumber\\
  &=&\int_{X'}\overline{\Theta}(M(x),R^{-1}(x'))
  \,d\mu(R^{-1}(x'))\nonumber\\
  &=&\int_{X'}\overline{\Theta}(M(x),R^{-1}(x'))
  \,d\nu(x')\nonumber\\
  &=&\int_{X'}\Theta(x,x')\,d\nu(x')\nonumber\\
  &=&F_{\nu}(x)\;.
\end{eqnarray}
Hence,
\begin{eqnarray}
  \int_{Y}F_{\mu}(y)\,\mathcal{D}_{\overline{\Theta},\overline{Z}}y
   &=&\int_{Y'}\overline{Z}(y')\,d\mu(y')\nonumber\\
  &=&\int_{X'}(\overline{Z}\circ R^{-1})(x')
  \,d\mu(R^{-1}(x'))\nonumber\\
  &=&\int_{X'}Z(x')
  \,d\nu(x')\nonumber\\
   &=&\int_{X}F_{\nu}(x)\,\mathcal{D}_{\Theta,Z}x\nonumber\\
   &=&\int_{X}F_{\mu}(M(x))\,\mathcal{D}_{\Theta,Z}x\;.
\end{eqnarray}$\QED$

It is important to keep in mind that
$\mathcal{D}_{\overline{\Theta},\overline{Z}}x$ and
$\mathcal{D}_{\Theta,Z}x$ are \emph{different} integrators.
However, when $M(X)=X$ and $M'$ is nuclear, the two are related by
Restriction \ref{theta relation}.
\begin{corollary}\label{cor. change of variable}
If $M(X)=X$ and $M'$ is nuclear and
$F_{\mu}(x)\in\mathcal{F}_R(X)$, then
\begin{equation}\label{Invariant integrator}
\int_{X}F_{\mu}(y)\,\mathcal{D}_{\overline{\Theta},\overline{Z}}y
   =\int_{X}F_{\mu}(M(x))|\mathrm{Det}M'_{(x)}|
   \,\mathcal{D}_{\overline{\Theta},\overline{Z}}x\;.
\end{equation}
\end{corollary}

\emph{Proof.} By (\ref{theta relation eq.}),
\begin{eqnarray}
   \int_{M(X)}F_{\mu}(y)\,\mathcal{D}_{\overline{\Theta},\overline{Z}}y
   &=&\int_{R^{-1}(X')}\overline{Z}(y')\,d\mu(y')\notag\\
  &=&\int_{X'}Z(x')\,d\nu(x')\notag\\
  &=&\int_{X'}|\mathrm{Det}R'_{(R^{-1}(x'))}|\overline{Z}(x')\,d\nu(x')\notag\\
  &=&\int_XF_{\nu}(x)|\mathrm{Det}M'_{(x)}|
  \,\mathcal{D}_{\overline{\Theta},\overline{Z}} x\notag\\
  &=&\int_XF_{\mu}(M(x))|\mathrm{Det}M'_{(x)}|
  \,\mathcal{D}_{\overline{\Theta},\overline{Z}} x\;.
\end{eqnarray}
The fourth line follows from the third by Definitions
\ref{def1}--\ref{def3}. $\QED$

Viewing
$F_{\mu}(y)\,\mathcal{D}_{\overline{\Theta},\overline{Z}}y$ as a
form, this result can be interpreted as
\begin{equation}
\int_{M(X)}F_{\mu}=\int_{X}M^{*}F_{\mu}\;.
\end{equation}
Loosely speaking, this says that the pull-back of the form in
`local coordinates' is
$(M^{*}F_{\mu})(x)=|\mathrm{Det}M'_{(x)}|F_{\mu}(M(x))$.
Alternatively, the corollary can be interpreted as a
transformation property of the integrator;
\begin{equation}
\mathcal{D}_{\overline{\Theta},\overline{Z}}(M(x))
=|\mathrm{Det}M'_{(x)}|
\,\mathcal{D}_{\overline{\Theta},\overline{Z}}x\;.
\end{equation}
Similar reasoning yields
\begin{equation}
\mathcal{D}_{\Theta,Z}(\ln M(x))=|\mathrm{Det}(\ln M)'_{(\ln x)}|
\,\mathcal{D}_{\Theta,Z}(\ln x)\;.
\end{equation}
In particular, for $M(x)=x+x_0$ where $x_0$ is a fixed element in
$X$ or $M(x)=e^{\epsilon \ln x_0} x$ with $\epsilon\in \R_+$, this
characterizes translation and scale invariant integrators on $\Za$
and $Z_a^+$ respectively.

\begin{proposition}[Integration by parts]\label{prop. integration by parts}
Suppose that $F_{\mu}\in \mathcal{F}_R^{\wedge}(X)$, then
\begin{equation}
\int_{X}\Bold{d}_y F_{\mu}(x) \;\mathcal{D}_{\Theta,Z}x
 =-\int_{X}F_{\mu}(x)\,\Bold{d}_y\mathcal{D}_{\Theta,Z}x\;.
\end{equation}
If $y$ is any fixed point $x_0\in X$, then
\begin{equation}\label{integration by parts}
0=\int_{X}\frac{\delta F_{\mu}(x)}{\delta x(\ti)}
\;\mathcal{D}_{\Theta,Z}x
 =\int_{X}F_{\mu}(x)\frac{\delta}{\delta
 x(\ti)}\mathcal{D}_{\Theta,Z}x\;.
\end{equation}
\end{proposition}
\emph{Proof.} The first equality follows trivially from
Restriction \ref{closed form}.

Now, if $x\mapsto x+hx_0$, then Restriction \ref{closed form} and
Proposition \ref{prop. Fubini} give
\begin{eqnarray}
  0&=&\int_{X}\Bold{d}_{x_0}[F_{\mu}(x)\mathcal{D}_{\Theta,Z}x]\notag\\
  &=&\int_{\Sigma}\int_{X}\left[\frac{\delta F_{\mu}(x)}{\delta x(\ti)}
\;\mathcal{D}_{\Theta,Z}x+F_{\mu}(x)\frac{\delta}{\delta
 x(\ti)}\mathcal{D}_{\Theta,Z}x\right]x_0(\ti)\,d\ti\;.
\end{eqnarray}
Then (\ref{integration by parts}) follows because $x_0(\ti)$ is an
arbitrary function and Corollary \ref{cor. change of variable}
implies $F_{\mu}(x+hx_0)=|\mathrm{Det}M'_{(x)}|^{-1}
F_{\mu}(x)=F_{\mu}(x)$ so that $\Bold{d}_{x_0}F_{\mu}(x)=0$.$\QED$

Equation (\ref{integration by parts}) can be viewed as an
infinitesimal characterization of an invariant integrator. For
example, if $X=\Za$ then it characterizes a translation invariant
integrator. Note, however, that it holds more generally; it is
true whenever $|\mathrm{Det}M'_{(z)}|=1$ for $M:\Za\rightarrow
\Za$. The same argument goes through for $z=\ln x$ with
$M(x)=e^{\epsilon \ln x_0}x$ yielding
\begin{equation}
0=\int_{\ln(X)}\frac{\delta F_{\mu}(z)}{\delta z(\ti)}
\;\mathcal{D}_{\Theta,Z}z
 =\int_{\ln(X)}F_{\mu}(z)\frac{\delta}{\delta
 z(\ti)}\mathcal{D}_{\Theta,Z}z\;.
\end{equation}

\section{Integrators}\label{app. integrators} This appendix
contains a discussion of the well-known Gaussian integrator, the
related Dirac integrator, and introduces two new integrators; the
Hermitian and gamma integrators. The Hermitian integrator is a
generalization of the Gaussian case, and it facilitates
integration of polynomials. The gamma integrator is an integrator
that is inspired by the Laplace transform of a gamma probability
distribution in much the same way as the Gaussian integrator is
inspired by the Fourier transform of a Gaussian probability
distribution.

\subsection{Gaussian and Dirac integrators}\label{app. gaussian
integrators} Let
\begin{eqnarray}
  &&\Theta(x,x';s)=\exp{\{-(\pi/s) Q(x)-2\pi i{\langle
  x',x\rangle}_{X}\}}\notag\\
  &&Z(x';s)=\exp{\{-\pi sW(x')\}}\;.
\end{eqnarray}
Here $s\in\C_+$, $Q$ is a nondegenerate bilinear form on
$X=H^1(\U\subseteq\M)$ such that $\mathrm{Re}\,(Q/s)>0$, and $W$
is its inverse on the dual space $X'$. Specifically,
$Q(x_1,x_2)=\langle Dx_1,x_2\rangle_{X}$ and $W(x'_1,x'_2)=\langle
x'_1,Gx'_2\rangle_{X}$ such that $DG=Id_{X'}$ and $GD=Id_{X}$. A
one parameter family of Gaussian integrators $\DQ x$ (or
$\mathcal{D}\omega_s(x)$) is defined according to the general
scheme by
\begin{equation}\label{gaussian}
  \int_{X}\exp^{\{-(\pi/s) Q(x)-2\pi i{\langle
  x',x\rangle}_{X}\}}\,\DQ x
  :=\int_{X}\exp^{\{-2\pi i{\langle
  x',x\rangle}_{X}\}}\,\mathcal{D}\omega_s(x):=\exp^{\{-\pi sW(x')\}}\;.
\end{equation}

The above characterization is for $\Theta:X\times X'\rightarrow\C$
and $Z:X'\rightarrow\C$. For the general case $\Theta:X\times
X'\rightarrow\C^d$ and $Z:X'\rightarrow\C^d$, (\ref{gaussian})
holds component-wise.

The Gaussian integrator is normalized,
\begin{equation}\label{gaussian normalization}
  \int_X\mathcal{D}\omega_s(x)=1\;;
\end{equation}
has zero mean,
\begin{equation}\label{zdero mean}
  \int_{X}\langle x',x\rangle_{X}\,\mathcal{D}\omega_s(x)
  =0\;;
\end{equation}
and covariance,
\begin{equation}\label{covariance}
  \int_{X}\langle x'_1,x\rangle_{X}\langle x'_2,x\rangle_{X}
  \,\mathcal{D}\omega_s(x)
  =\frac{s}{2\pi}W(x'_1,x'_2)\;.
\end{equation}
In terms of coordinate functions, the covariance can be
conveniently written as
\begin{equation}
  \int_{X}x^{\alpha}(\ti)x^{\beta}(\mathrm{u})
  \,\mathcal{D}\omega_s(x)
  =\frac{s}{2\pi}G^{\alpha\beta}(\ti,\mathrm{u})\;;
\end{equation}
where $G^{\alpha\beta}(\ti,\mathrm{u})$ is the (nondegenerate)
Green's function of $D$, i.e., for $x'_1=\delta^{\alpha}_{\ti}$
and $x'_2=\delta^{\beta}_{\mathrm{u}}$,
$W(\delta^{\alpha}_{\ti},\delta^{\beta}_{\mathrm{u}})
=\langle\delta^{\alpha}_{\ti},G\delta^{\beta}_{\mathrm{u}}\rangle
=:G^{\alpha\beta}(\ti,\mathrm{u})$ with
$\alpha,\beta\in\{1,\ldots,m\}$. The elements of $X$ are pointed
paths, and so $G^{\alpha\beta}(\ti,\mathrm{u})$ inherits boundary
conditions from the paths.

The change of variable formula (\ref{change of variable}) can be
used to reduce integrals of the type $\int_Xf(\langle
x_1',x\rangle)\mathcal{D}\omega_s (x)$ to finite dimensional
integrals:
\begin{proposition}\label{prop. gaussian reduction}
Let $L:X\rightarrow\R^n$ by $x\mapsto
\mathbf{u}=\{\mathrm{u}^1,\ldots,\mathrm{u}^n\}$ where
$\mathrm{u}^i=\langle {x'}_i,x\rangle_X$ for a given ${x'}_i\in
X'$ and denote $W_{\R^m}=(W\circ\widetilde{L})$ so that
$Q_{\R^m}(\mathbf{u})= [W_{\R^m}^{-1}]_{ij}\mathrm{u}^i
\mathrm{u}^j:=W_{\R^m}^{-1}(\mathbf{u})$, then
\begin{equation}\label{gaussian reduction}
\int_Xf(\langle x_1',x\rangle)\mathcal{D}\omega_s (x)
=\int_{\R^n}f(\mathbf{u})|\mathrm{det}\,sW_{ij}|^{-1/2}\exp^{\{-(\pi
s)W^{-1}_{ij}\mathrm{u}^i\mathrm{u}^j-2\pi
i\langle\mathbf{u}',\mathbf{u}\rangle_{\R^n}\}}\,d\mathbf{u}\;.
\end{equation}
\end{proposition}

\emph{Proof.} The proof is straightforward using (\ref{change of
variable}). The determinant factor is a consequence of the
normalization (\ref{gaussian normalization}).$\QED$ More
generally, integrals of the form
\begin{equation}
\int_Xf(\langle x_1',x\rangle,\ldots,\langle x_n',x\rangle
)\mathcal{D}\omega_s (x)
\end{equation}
reduce to finite dimensional integrals and can ultimately be
evaluated in terms of $G$ as is well known.

Gaussian integrators possess some important properties that enable
the construction of path integral solutions to second order
partial differential equations. The first is the semi-group
property (\cite{CA/D-M}).

\begin{proposition}\label{prop. semi-group}
Let
$(U_{\ti}F)(\x_a):=\int_{\Za}F(\x_a\cdot\mathit{\Sigma}(\ti,z))
\,\mathcal{D}\omega_s(z)$, then
\begin{eqnarray}\label{semi group}
  (U_{\ti_c}(U_{\ti_b}F))(\x_a)
  &=&(U_{\ti_c+\ti_b}F)(\x_a)\;
\end{eqnarray}
where $\ti_c>\ti_b>0$.
\end{proposition}

\emph{Proof.} Define the path
$z:[0,\tr_b+\tr_c]\rightarrow\R^d(\C^d)$ by
\begin{equation}
z(\tr)=\left\{\begin{array}{ll}
  z_1(\tr)\;\;\;\;\mbox{for}\;\;0\leq \tr\leq \tr_b\\
  z_1(\tr_b)+z_2(\tr-\tr_b)\;\;\;\;\mbox{for}
  \;\;\tr_b\leq \tr\leq \tr_b+\tr_c
  \end{array}\right.
\end{equation}
where $z_1\in \Za_1$ and $z_2\in \Za_2$. Then
\begin{eqnarray}
(U_{\tr_c}(U_{\tr_b}F))(\x_a) &=&\int_{\Za_2}(U_{\tr_b}F)
(\x_a\cdot\mathit{\Sigma}(\tr_c,z_2))\,\mathcal{D}\omega_s(z_2)\nonumber\\
  &=&\int_{\Za_2}\int_{\Za_1}
  F(\x_a\cdot\mathit{\Sigma}(\tr_c,z_2)
  \cdot\mathit{\Sigma}(\tr_b,z_1))
  \,\mathcal{D}\omega_s(z_1)\mathcal{D}\omega_s(z_2)\;.\notag\\
\end{eqnarray}
By the uniqueness of the solution to the differential equation
(\ref{tau parametrization}) and the Fubini relation
(\ref{Fubini}), this becomes
\begin{eqnarray}\label{semi-group}
(U_{\tr_c}(U_{\tr_b}F))(\x_a) &=&\int_{\Za_2}\int_{\Za_1}
  F(\x_a\cdot\mathit{\Sigma}(\tr_c+\tr_b,(z_1,z_2)))
    \,\mathcal{D}\omega_s(z_1)\mathcal{D}\omega_s(z_2)\nonumber\\
    &=&\int_{\Za}F(\x_a\cdot\mathit{\Sigma}(\tr_c+\tr_b,z))
    \,\mathcal{D}\omega_s(z)\nonumber\\
    &=&(U_{\tr_c+\tr_b}F)(\x_a)\;.
\end{eqnarray}$\QED$

Next, under an affine map $A:X\rightarrow \widetilde{X}$ by
$x\mapsto Ax=Lx+x_0$ where $L$ is an invertible linear map and
$x_0\in \widetilde{X}$ is a fixed element, the image of $\DQ{x}$
is
\begin{equation}\label{integrator image}
  \DQ x=|\mathrm{Det}L|^{-1}\mathcal{D}_{Q,W}A(x)\;.
\end{equation}
This follows from Corollary \ref{cor. change of variable}. In
particular, for $L\equiv Id$, (\ref{integrator image}) expresses
the fact that a gaussian integrator is translation invariant.
Finally, for $F(x)\in \mathcal{F}_R(X)$, gaussian integrators
satisfy the integration by parts formula,
\begin{equation}
 \int_{X}\frac{\delta F(x)}{\delta
  x(\ti)}\;\DQ{x}=-\int_{X}F(x)\frac{\delta}{\delta
  x(\ti)}\DQ{x}=0\;.
\end{equation}
This is just the infinitesimal form of translation invariance of
the integrator.

Related to Gaussian integrators is the Dirac integrator
$\mathcal{D}\delta(x)$. The definition of the Dirac integrator
(\cite{LA}) is based on the fact that a delta function can be
represented by a Gaussian with zero width. It is characterized by
\begin{eqnarray}
  \lim_{|s|\rightarrow 0}\int_{X}\exp{\{-(\pi/s) Q(x)-2\pi i{\langle
  x',x\rangle}_{X}\}}\DQ x\notag\\
  &&\hspace{-1in}=:\int_{X}\exp{\{-2\pi i{\langle
  x',x\rangle}_{X}\}}\delta_X(x)\DQ x\notag\\
  &&\hspace{-1in}=:\int_{X}\exp{\{-2\pi i{\langle
  x',x\rangle}_{X}\}}\mathcal{D}\delta(x)\notag\\
  &&\hspace{-1in}=\lim_{|s|\rightarrow 0}\exp{\{-\pi sW(x')\}}\notag\\
  &&\hspace{-1in}=1\;.
\end{eqnarray}
Conversely, the limit $|s|\rightarrow\infty$ defines the inverse
Dirac integrator
 \begin{eqnarray}
 \lim_{|s|\rightarrow \infty}\int_{X}\exp{\{-(\pi/s) Q(x)-2\pi i{\langle
  x',x\rangle}_{X}\}}\DQ x\notag\\
  &&\hspace{-1in}=:\int_{X}\exp{\{-2\pi i{\langle
  x',x\rangle}_{X}\}}\mathcal{D}\delta^{-1}(x)\notag\\
  &&\hspace{-1in}=\lim_{|s|\rightarrow \infty}\exp{\{-\pi sW(x')\}}\notag\\
  &&\hspace{-1in}=:\delta_{X'}(x')\;.
\end{eqnarray}
In short hand, $\lim_{|s|\rightarrow
0}\mathcal{D}\omega_s(x)=:\mathcal{D}\delta(x)$ and
$\lim_{|s|\rightarrow
\infty}\mathcal{D}\omega_s(x)=:\mathcal{D}\delta^{-1}(x)$.

This definition is `good' because, under the linear map
$L:X\rightarrow\R^n$ by $x\mapsto
\mathbf{u}=\{\mathrm{u}^1,\ldots,\mathrm{u}^n\}$ where
$\mathrm{u}^i=\langle {x'}_i,x\rangle_X$ for a given ${x'}_i\in
X'$, the integral reduces (using Proposition \ref{prop. gaussian
reduction}) as expected for any $n$;
\begin{eqnarray}
\int_{X}\exp{\{-2\pi i{\langle
  x',x\rangle}_{X}\}}\mathcal{D}\delta(x)
  &\rightarrow&\lim_{|s|\rightarrow
  0}\int_{\R^n}\exp\{-(\pi s)W_{\R^n}^{-1}\mathbf{u}
  -2\pi i\langle\mathbf{u}',\mathbf{u}\rangle_{\R^n}\}\notag\\
  &&\hspace{.6in}
  |\mathrm{det}\,sW_{\R^n}|^{-1/2}d\mathbf{u}\notag\\
  &=&\int_{\R^n}\exp\{-2\pi i\langle
 \mathbf{u}',\mathbf{u}\rangle_{\R^n}\}
 \delta^n(\mathbf{u})d\mathbf{u}\notag\\
  &=&1\;.
\end{eqnarray}
Similarly,
\begin{eqnarray}
\int_{X}\exp{\{-2\pi i{\langle
  x',x\rangle}_{X}\}}\mathcal{D}\delta^{-1}(x)
  &\rightarrow&\lim_{|s|\rightarrow
  \infty}\int_{\R^n}\exp\{-(\pi s)W_{\R^n}^{-1}\mathbf{u}
  -2\pi i\langle\mathbf{u}',\mathbf{u}\rangle_{\R^n}\}\notag\\
  &&\hspace{.6in}
  |\mathrm{det}\,sW_{\R^n}|^{-1/2}d\mathbf{u}\notag\\
  &=&\int_{\R^n}\exp\{-2\pi i\langle
 \mathbf{u}',\mathbf{u}\rangle_{\R^n}\}d\mathbf{u}\notag\\
  &=&\delta^n(\mathbf{u}')\;.
\end{eqnarray}

The Dirac integrator possesses the expected properties;
\begin{proposition}
\begin{equation}
\int_{X}F_{\mu}(x)\mathcal{D}\delta(x)=F_{\mu}(0)\;,
\end{equation}
and for non-trivial arguments,
\begin{equation}
\int_{X}F_{\mu}(x)\mathcal{D}\delta(M(x))
=\sum_{x_o}|\mathrm{Det}M_{(x_o)}'|^{-1}F_{\mu}(x_o)\;.
\end{equation}
\end{proposition}

\emph{Proof.} First,
\begin{eqnarray}
\int_{X}F_{\mu}(x)\mathcal{D}\delta(x)
&=&\int_{X}\delta(x)\int_{X'}\exp{\{-(\pi/s)Q(x)-2\pi i{\langle
  x',x\rangle}_{X}\}}d\mu(x')\,\DQ x\notag\\
&=&\lim_{|s|\rightarrow 0}
\int_{X'}\int_{X}\exp{\{-(2\pi/s)Q(x)-2\pi i{\langle
  x',x\rangle}_{X}\}}\DQ x\,d\mu(x')\notag\\
&=&\lim_{|s|\rightarrow 0} \int_{X'}\exp{\{-(\pi s/2)W(x')\}}\,d\mu(x')\notag\\
&=&\int_{X'}d\mu(x')\notag\\
&=&F_{\mu}(0)\;,
\end{eqnarray}
and second,
\begin{eqnarray}
\int_{X}F_{\mu}(x)\mathcal{D}\delta(M(x))
&=&\int_{X}F_{\mu}(x)\delta(M(x))\DQ x\notag\\
&=&\int_{Y}F_{\mu}(M^{-1}(y))\delta(y)
\mathcal{D}_{\overline{Q},\overline{W}}(M^{-1}(y))\notag\\
&=&\int_{Y}F_{\mu}(M^{-1}(y))\delta(y)|\mathrm{Det}{M_{(y)}^{-1}}'|
\mathcal{D}_{\overline{Q},\overline{W}}(y)\notag\\
&=&\sum_{x_o}|\mathrm{Det}M_{(x_o)}'|^{-1}F_{\mu}(x_o)
\end{eqnarray}
where $M(x_o)=0$. $\QED$

In short-hand notation,
$\delta(M(x))=\sum_{x_o}|\mathrm{Det}M_{(x_o)}'|^{-1}\delta(x-x_o)$.

\subsection{Hermite integrator}
Another integrator of particular interest is a generalization of
the Gaussian integrator. For this application, the space of paths
is $\Za=H^1(\U\subseteq\M)$. $\Za$ is equipped with a quadratic
form $Q(z)=\langle Dz,z\rangle_{\Za}$ and the dual space $\Za'$ is
equipped with a related quadratic form $W(z')=\langle
z',Gz'\rangle_{\Za'}$ such that $DG=Id_{\Za'}$ and $GD=Id_{\Za}$.
Assume $W$ is nuclear and $F_{\mu}(x)\in\mathcal{F}_R(X)$.

Let
\begin{eqnarray}
  \Theta_n(z,z';s)&:=&\left|\mathrm{Det}\,\frac{\pi s
W}{2}\right|^{n/2}
    H_n\left(\sqrt{\pi Q(z)/s}\right)\exp^{\{-\pi
  Q(z)/s-2\pi i\zdual\}}\nonumber\\
  &=:&\left|\mathrm{Det}\,\frac{\pi s
W}{2}\right|^{n/2}
    H_n(\hat{z})\exp^{\{-\hat{z}^2-2\pi i\zdual\}}\nonumber\\
  &=:&
  \widehat{H}_n(\hat{z};s)\exp^{\{-\hat{z}^2-2\pi i\zdual\}}
\end{eqnarray}
with $s\in\C_+$ and $H_n(\hat{z})$ the $n$-th order functional
Hermite polynomial. If the space $\Za$ has the structure
$Z=Z_1\cup Z_2\cdots\cup Z_r$ such that $Z_1\cap Z_2\cdots\cap
Z_r=\emptyset$, require
\begin{equation}
  \widehat{H}_n(\hat{z}_1,\hat{z}_2,\ldots,\hat{z}_r;s)\equiv
\widehat{H}_{n}(\hat{z}_1;s)\widehat{H}_{n}
(\hat{z}_2;s)\times\ldots\times \widehat{H}_{n}(\hat{z}_r;s)\;.
\end{equation}

Let
\begin{equation}\label{Hermite Z}
  Z_n(z';s):=
  \int_{\Za'}\widehat{H}_{n'}\left(\sqrt{\pi s W(y')};s\right)
  \exp^{-\pi s W(z'-y')}
  \;d\mu (y')
\end{equation}
where
\begin{equation}
  \left.\widehat{H}_{n'}(z';s)
  :=\widehat{H}_n(z;s)\right|_{z^j\rightarrow\delta^{(j)}(z')}\;.
\end{equation}
Then the $n$-th order Hermite integrator is characterized by
\begin{equation}\label{Hermite integrator}
\int_{\Za}\exp^{\left\{-2\pi i\zdual_{\Za}\right\}}
\mathcal{D}\rho_{n,s}(z):=\int_{\Za}\widehat{H}_n(\hat{z};s)
\exp^{\left\{-\hat{z}^2 -2\pi i
\zdual_{\Za}\right\}}\DQ{z}=Z_n(z';s)\;.
\end{equation}
It follows from (\ref{Hermite Z}) and (\ref{Hermite integrator})
that the Hermite integrator is normalized according to
\begin{equation}\label{Hermitian normalization}
  \int_{\Za}\,\mathcal{D}\rho_{n,s}(z)
  =\left\{\begin{array}{cl}
  1&\mbox{for}\,n=0\\
  0&\mbox{otherwise}
  \end{array}\right.\;.
\end{equation}
Clearly, $\mathcal{D}\rho_{0,s}({z})$ is the familiar Gaussian
integrator. Integrable functionals are of the form
\begin{equation}
  F_{\mu}(z;n,s)=H_n(\hat{z})\exp^{-\hat{z}^2}\int_{\Za'}
  \exp^{\left\{-2\pi i\zdual_{{\Za}}\right\}}\,d\mu({z}')\;.
\end{equation}

Under the linear map $L:\Za\rightarrow\R$ by
$z\mapsto\zdual\in\R$, the quadratic form becomes
\begin{equation}
  (Q\circ L^{-1})(\zdual)=:Q_{\R}(\zdual)
  =\zdual^2/W({z}')\;.
\end{equation}
Hence, Proposition \ref{prop. gaussian reduction} yields
\begin{eqnarray}\label{1-dim Hermite}
\int_{\Za}\,\mathcal{D}\rho_{n,s}(z) &=&\left|\frac{\pi s
W(z')}{2}\right|^{n/2}|sW({z}')|^{-1/2}\nonumber\\
&&\hspace{.5in}\times
\int_{\R}H_n\left(\begin{array}{l}\sqrt{\frac{\pi}{sW({z}')}}\end{array}
\mathrm{u}\right)
\exp^{\left\{-\pi\mathrm{u}^2/sW({z}')\right\}}\,d\mathrm{u}\;.
\end{eqnarray}

More generally, let $L:\Za\rightarrow\R^m$ by $z\mapsto
\mathbf{u}$ and denote $W_{\R^m}=(W\circ\widetilde{L})$, then
$Q_{\R^m}(\mathbf{u})= [W_{\R^m}^{-1}]_{ij}\mathrm{u}^i
\mathrm{u}^j:=W_{\R^m}^{-1}(\mathbf{u})$ and
\begin{eqnarray}\label{n-dim Hermite}
  \int_{\Za}f(\zdual)\;\mathcal{D}\rho_{n,s}(z)
&=&\left|\mathrm{det}\,\frac{\pi s
W_{\R^m}}{2}\right|^{n/2}|\mathrm{det}\,sW_{\R^m}|^{-1/2}\nonumber\\
&&\hspace{.0in}\times\int_{\R^m}f(\mathbf{u})
H_n\left(\begin{array}{l} \sqrt{\frac{\pi
W_{\R^m}^{-1}(\mathbf{u})}{s}}\end{array} \right)
\exp^{\{-(\pi/s)W_{\R^m}^{-1}(\mathbf{u})\}}\;d\mathbf{u}\nonumber\\
\end{eqnarray}
where $H_n(\mathbf{u}):=\Pi_{i=1}^{m}H_{\alpha_{i}}(\mathrm{u}^i)$
with $|\alpha|:=\Sigma_{i=1}^m(\alpha_i)=n$ is an $m$-fold product
of Hermite polynomials (see e.g. \cite{DU/XU}).

The Hermite integrator facilitates integration of normal ordered
(Wick ordered) monomials:
\begin{proposition} Define the normal ordered functional monomial $:
\hspace{-.05in}z^m\hspace{-.05in}:$ by $:
\hspace{-.05in}z^m\hspace{-.05in}:\;=\left|\mathrm{Det}\,\frac{\pi
s W}{2}\right|^{m/2}H_m(\sqrt{\pi Q(z)/s})$. Let
$L:\Za\rightarrow\R$ by $z\mapsto\zdual$, then
\begin{equation}
\int_{\Za}:\zdual^m:\;\mathcal{D}\rho_{n,s}(z) =|\pi
sW(z')|^mn!\,\delta_{nm}
\end{equation}
where $:\zdual^m:$ is a normal ordered monomial.
\end{proposition}

\emph{Proof.} Under $L:\Za\rightarrow\R$,
\begin{equation}
:z^m:\;\mapsto \;:\zdual^m:\;=\left|\frac{\pi s
W(z')}{2}\right|^{m/2}H_m\left(\begin{array}{l}\sqrt{\frac{\pi}{sW({z}')}}\end{array}
\zdual\right)\;.
\end{equation}
The integral follows immediately from (\ref{1-dim Hermite}) and
the definition of $:z^m:$. In a Fock representation $|n\rangle$
for the simple harmonic oscillator with ground state $|0\rangle$,
\begin{equation}
\zdual^m=\left|\frac{\pi s W(z')}{2}\right|^{m/2}(a+a^{\dag})^m\;.
\end{equation}
The binomial expansion and normal ordering yield (see e.g.
\cite{GL/JA})
\begin{equation}
 :\zdual^m:|n\rangle=\left|\frac{\pi s W(z')}{2}\right|^{m/2}
 \sum_{i=0}^m\left(
 \begin{array}{c}m\\i\end{array}\right)a^{\dag \,i}a^{(m-i)}\;|n\rangle\;.
 \end{equation} $\QED$
\begin{corollary}
\begin{equation}
\sum_{n}\frac{1}{n!}\int_{\Za}:e^{\zdual}:\,\mathcal{D}\rho_{n,s}(z)
=e^{\pi sW(z')}\;.
\end{equation}
\end{corollary}

\subsection{Gamma integrator}\label{app. gamma integrator}
Specialize to the case where the space of paths $\zTa$ is the
Sobolev space $H^1(\Sigma)$, i.e. $z\in\zTa=H^1(\Sigma)$. In order
to more clearly differentiate between elements of $\Za$ and
$\zTa$, it is convenient to now switch notation and write
$\tr\in\Ta\equiv\zTa$. (Recall that $\tr=\ln x$ with $x\in
Z_a^1$.) $\Ta$ is a Hilbert space with Hermitian inner product
\begin{equation}
  (\tr_1,\tr_2)_{H^1}=\int_{\Sigma}\dot{\tr_1}\dot{{\tr}_2}^*\;d\ti
  \hspace{.5in}\tr_1,\tr_2\in\Ta\;.
\end{equation}
The range of $\tr$ is taken to be the path
$C_+:=\tr(\Sigma)=[0,\tr(\ti_b)]\cup[0,{\tr}^*(\ti_b)]\subset\C_+$
with an associated direction determined by the order of the
intervals.

The dual space $\T$ is the space of linear forms
$\tr':\Ta\rightarrow\C$ by
\begin{equation}
  \tr\mapsto\dual_{\Ta}:=\int_{\Sigma}\tr'(\ti)\tr(\ti)d\ti\;.
\end{equation}
In particular, define the linear form $T_{\omega}:\Ta\rightarrow
\C$ by
\begin{equation}
  \tr\mapsto
  \omega\int_{\Sigma}\dot{\tr}(\ti)d\ti\;,
\end{equation}
where $\omega\in\C$. $\T$ is equipped with complex Borel measures
$\mu$.

Using the general scheme, characterize a two parameter family of
integrators by
\begin{equation}\begin{array}{l}
  \Theta(\tr,\tr';\omega)=\exp\{T_{\omega}(\tr)-\dual_{\Ta}\}\\
  \\
  Z(\tr';\omega,\nu)=(\overline{\tr'}-\omega)^{-\nu}\;\;;
  \hspace{.2in}\mathrm{Re}(\overline{\tr'}-\omega)>0
  \end{array}
\end{equation}
where $\tr\in\Ta$ is the path $\tr:\Sigma\rightarrow \C_+$,
$\tr'\in \T$, $\nu\in\C_+$, and
\begin{equation}
  \overline{\tr'}:=\frac{\int_{\Sigma}\tr'\dot{\tr}\;d\ti}
  {\int_{\Sigma}\dot{\tr}\;d\ti}
  =\frac{\int_{\tr(\Sigma)}\tr'\;d\mathrm{s}}{\tr(\Sigma)}\;.
\end{equation}

The gamma integrator $\mathcal{D}_{\nu}\tr$ (or
$\mathcal{D}\gamma_{\omega,\nu}(\tr)$) is defined by
\begin{eqnarray}\label{gamma integrator}
  \int_{\Ta}\exp\{T_{\omega}(\tr)-\dual_{\Ta}\}\,\Da{\tr}
  :=\int_{\Ta}\exp\{-\dual_{\Ta}\}\,\mathcal{D}\gamma_{\omega,\nu}(\tr)
 :=\frac{1}{(\overline{\tr'}-\omega)^{\nu}}\,.\nonumber\\
\end{eqnarray}
Note the normalization
\begin{equation}\label{gamma normalization}
\int_{\Ta}\,\mathcal{D}\gamma_{\omega,\nu}(\tr)=(-\omega)^{-\nu}\;.
\end{equation}
 The associated space of integrable
functionals $\mathcal{F}_R(\Ta)$ contains elements of the form
\begin{equation}
  F_{\mu}(\tr;\omega):=\exp\{T_{\omega}(\tr)\}
  \int_{\T}\exp\{-\dual_{\Ta}\}\;d\mu(\tr')\;.
\end{equation}
On $\mathcal{F}_R(\Ta)$, the integral operator
$\int_{\Ta}\mathcal{D}_{\nu}\tr$ is defined by
\begin{equation}
 \int_{\Ta}F_{\mu}(\tr;\omega)\mathcal{D}_{\nu}\tr
 :=\int_{\T}(\overline{\tr'}-\omega)^{-\nu}d\mu(\tr')\;.
\end{equation}
According to Corollary \ref{cor. change of variable} and the
characterization (\ref{gamma integrator}), $\mathcal{D}_{\nu}\tr$
is scale invariant; i.e., in short-hand notation
$\mathcal{D}_{\nu}(\epsilon\tr)=\mathcal{D}_{\nu}\tr$. This can be
seen directly from (\ref{gamma integrator}), which also implies
$\mathcal{D}\gamma_{\omega,\nu}(\epsilon\tr)
=\mathcal{D}\gamma_{\epsilon\omega,\nu}(\tr)$.

Typical applications lead to integrals of the type
\begin{equation}
\int_{\Ta}F_{\mu}(\tr(\ti_b);\omega)\mathcal{D}_{\nu}\tr\;.
\end{equation}
where $\tr(\ti_b)$ is usually either pure real or pure imaginary.
The change of variables formula (\ref{change of variable}) allows
this type of integral to be reduced to a one dimensional integral.
Consider the special case where
$\Sigma=\mathbb{T}:=[\ti_a,\ti_b]\subseteq\R$, $-\omega\in\R_+$,
and $C_+=\R_+$. Then, under the linear map $L:\Ta\rightarrow\R_+$,
the change of variable formula (\ref{change of variable}) and the
definition (\ref{gamma integrator}) give,
\begin{equation}
  \int_{\Ta}\mathcal{D}\gamma_{\omega,\nu}(\tr)
  =C\int_{\R_+}
  \exp\{\omega \mathrm{u}\}\mathrm{u}^{\nu-1}\,d\mathrm{u}
\end{equation}
where the constant $C$ is determined by the normalization
condition (\ref{gamma normalization}) to be $\Gamma(\nu)^{-1}$.
Hence,
\begin{proposition}
Under the linear map $L:\Ta\rightarrow\R_+$ the gamma integrator
reduces to
\begin{equation}\label{reduced gamma}
  \int_{\Ta}\,\mathcal{D}\gamma_{\omega,\nu}(\tr)
  \rightarrow\int_{\R_+} \frac{\exp\{\omega \mathrm{u}\}}{\Gamma(\nu)}\,
  \mathrm{u}^{\nu}\,d(\ln\mathrm{u})\;.
\end{equation}
\end{proposition}
Up to a factor of $(-\omega)^\nu$, the integrand of (\ref{reduced
gamma}) is just the gamma probability distribution which justifies
the appellation `gamma' integrator.

\begin{corollary}
When $\tr(\ti)$ is pure imaginary, the linear map
$L:\Ta\rightarrow i\R$ induces the reduction
\begin{eqnarray}
  \int_{\Ta}\,\mathcal{D}\gamma_{\omega,\nu}(\tr)
  &\rightarrow &
  \frac{1}{2}\int_{0}^{i\infty}\frac{\exp\{\omega \mathrm{u}\}}{\Gamma(\nu)}\,
  \mathrm{u}^{\nu}\,d(\ln\mathrm{u})
  +\frac{1}{2}\int_{0}^{-i\infty}\frac{\exp\{\omega \mathrm{u^*}\}}{\Gamma(\nu)}\,
  \mathrm{u^*}^{\nu}\,d(\ln\mathrm{u^*})\nonumber\\
  &=&\frac{1}{2}\int_{i\R}\frac{\exp\{\omega \mathrm{u}\}}{\Gamma(\nu)}\,
  \mathrm{u}^{\nu}\,d(\ln\mathrm{u})
\end{eqnarray}
where $\mathrm{u}=-{\mathrm{u}^*}$.
\end{corollary}

As another example, let $\Sigma=\R_+$ and $\tr(\Sigma)=S^1$ with
$F_{\mu}$ periodic $F_{\mu}(\tr(\ti))=F_{\mu}(\tr(\ti+2\pi))$.
\begin{proposition}
Under the map $L:\Ta\rightarrow S^1$; the gamma integrator reduces
to
\begin{equation}
\int_{\Ta}\,\mathcal{D}\gamma_{\omega,\nu}(\tr)
 \rightarrow
 (1-e^{\omega})
 \int_{0}^1  \frac{\exp\{\omega \mathrm{u}\}}{\gamma(\nu,-\omega)}\,
  \mathrm{u}^{\nu-1}\,d\mathrm{u}
\end{equation}
where $\gamma(\nu,-\omega)=\nu^{-1}(-\omega)^{\nu}
\;_1F_1(\nu,\nu+1;\omega)$.
\end{proposition}

\emph{Proof.} The proof follows the same reasoning as Proposition
\ref{prop. gaussian reduction}. The periodicity of $F_{\mu}$
accounts for the different domain of integration and the
normalization terms.$\QED$

\begin{proposition}
The gamma integrator satisfies
\begin{equation}
\int_{\Ta}\tr(\ti_b)^{\varrho}\,\mathcal{D}\gamma_{\omega,\nu}(\tr)
  =\frac{\Gamma(\nu+\varrho)}{\Gamma(\nu)}
  \int_{\Ta}\,\mathcal{D}\gamma_{\omega,(\nu +\varrho)}(\tr)
\end{equation}
for $\varrho\in \C_+$ and $\tr(\Sigma)=C_+$. If $\tr(\Sigma)=S^1$
then the constant factor on the right-hand side is
$\gamma(\nu+\varrho,-\omega)/\gamma(\nu,-\omega)$.
\end{proposition}

\emph{Proof.} The change of variable $\tr\mapsto
\tr(\ti_b)=\mathrm{u}\in C_+$ yields
\begin{eqnarray}
\int_{\Ta}\tr(\ti_b)^{\varrho}\,\mathcal{D}\gamma_{\omega,\nu}(\tr)
&=&\int_{C_+} \frac{\exp\{\omega \mathrm{u}\}}{\Gamma(\nu)}\,
  \mathrm{u}^{\nu+\varrho}\,d(\ln\mathrm{u})\notag\\
  &=&\frac{\Gamma(\nu+\varrho)}{\Gamma(\nu)}
  \int_{C_+} \frac{\exp\{\omega \mathrm{u}\}}{\Gamma(\nu+\varrho)}\,
  \mathrm{u}^{\nu+\varrho}\,d(\ln\mathrm{u})\notag\\
  &=&\frac{\Gamma(\nu+\varrho)}{\Gamma(\nu)}(-\omega)^{_(\nu+\varrho)}\notag\\
  &=&\frac{\Gamma(\nu+\varrho)}{\Gamma(\nu)}
  \int_{\Ta}\,\mathcal{D}\gamma_{\omega,(\nu +\varrho)}(\tr)\;.
\end{eqnarray}
$\QED$

For $\varrho$ a real integer this gives the moments of the gamma
integrator, but note that it holds for $\varrho\in \C_+$ in
general.

\begin{proposition}
Under scaling with respect to $\omega$,
\begin{equation}
\int_{\Ta}\,\mathcal{D}\gamma_{\epsilon\omega,\nu}(\tr)
=\epsilon^{-\nu}\int_{\Ta}\,\mathcal{D}\gamma_{\omega,\nu}(\tr)\;.
\end{equation}
\end{proposition}

\emph{Proof.} The equality follows trivially from the
characterization (\ref{gamma integrator}). $\QED$

A particular element of the family
$\mathcal{D}\gamma_{\omega,\nu}$ that will be useful later is
defined by
\begin{equation}\label{d-tau}
   \int_{\Ta}F(\tr)\DO{\tr}
  :=\lim_{\stackrel{\scriptstyle\nu\rightarrow
  0^+}{|\omega |\rightarrow\infty}}
  \int_{\Ta}F(\tr)\mathcal{D}\gamma_{\omega,\nu}(\tr)\;.
\end{equation}

Hence, for integrands of the form $F(\dual)$,
\begin{equation}\label{reduced tau}
  \int_{\Ta}F(\dual)\,\DO{\tr}
  =\mathcal{N}\int_{C_+} F(\ti)\,d(\ln \ti)
\end{equation}
where $C_+\subset\C_+$ is the range of $\tr$. The normalization
factor $\mathcal{N}$ replaces the indeterminate factor
$\exp^{\infty}/\Gamma(0^+)$ and is fixed by a suitable
normalization condition on $F(\ti)$.

\begin{proposition}\label{scale invariance}
$\DO{\tr}$ is scale invariant and normalized.
\end{proposition}

\emph{Proof.} Scale invariance follows immediately from
(\ref{gamma integrator}) and (\ref{d-tau}). The normalization is
\begin{equation}\label{tau normalization}
\int_{\Ta}\,\DO{\tr}
  =\lim_{\stackrel{\scriptstyle\nu\rightarrow
  0^+}{|\omega |\rightarrow\infty}}
  \int_{\Ta}\mathcal{D}\gamma_{\omega,\nu}(\tr)
  =\lim_{\stackrel{\scriptstyle\nu\rightarrow
  0^+}{|\omega |\rightarrow\infty}}(-\omega)^{-\nu}
  =1\;.
\end{equation}
$\QED$

\section{Variational Principle}\label{sec. variation}

A key step in constructing path integral solutions to PDEs with
finite boundaries is to introduce a time reparametrization denoted
by $\tau$. The first exit time\footnote{The term ``time'' is used
here to refer to the evolution parameter of some path. Though
expedient and conventional, its use can be problematic in some
situations; I will be more careful when warranted.} mentioned in
the introduction, which is related to $\tau$, is well motivated
from a stochastic point of view. The concept of first exit time
and Ito's formula lead readily to the functional integrals of
interest in the stochastic formulation. Unfortunately, it is not
immediately obvious how to translate these constructions into a
path integral formulation. Moreover, since the path integral
formulation would be expected to include the stochastic results as
a subset, it would be advantageous to develop the path integral
constructions based on concepts basic to the path integral
formulation rather than to simply translate stochastic concepts.
Therefore, it is instructive to formulate a physical motivation
for introducing the parameter $\tr$. In doing so, some insight and
physical intuition is gained for the solutions that are
constructed.

First, recall Feynman's reasoning (\cite{FE}) that produced a path
integral solution of the Schr\"{o}dinger-type parabolic PDE.
Roughly speaking, the prescription is to: choose a wave function
that represents the state of a physical system at a given point at
a given time; weight this wave function by the exponentiated
classical action functional for paths with \emph{fixed} initial
and final positions; and ``integrate'' over all such paths. From
this prescription, it is not immediately obvious how to proceed
when the physical system is contained in a bounded region (See
however \cite{GR1}, \cite{GR2}, \cite{GR3} and references therein
for an alternative approach.).

A hint about how to proceed appears when one constructs a path
integral representation of a fixed-energy transition amplitude
(\cite{LA}). The variational principle for paths with fixed energy
differs from the case of unrestricted paths in that a time
reparametrization is required (\cite{AB/MA}). The fixed energy
path integral yields a solution of the elliptic Dirichlet problem,
and the time reparametrization can be re-interpreted as a path
reparametrization such that all reparametrized paths which start
at a given point at an initial time reach some \emph{boundary}
(not some point) at a final time.

Now, some of the paths of interest for a physical system with a
boundary are paths that have a fixed initial point and intersect a
given boundary at some time. This description corresponds to a
variational problem from a fixed initial point to a
\emph{manifold} in the dependent--independent variable space (see
e.g. \cite{SA}). It turns out that this type of variational
problem incorporates a variable endpoint in the functional
integral which can be interpreted as parametrizing the time it
takes a path to reach some boundary, i.e. a first exit time.
Alternatively, the variable endpoint can be interpreted as a
non-dynamical dependent variable which takes into account the
implicit constraint induced by a physical system with a boundary.

To formulate the variational principle for paths taking their
values in a manifold $\M$ that intersect a boundary, consider the
$\mathrm{dim}(m+1)$ dependent--independent variable space
$\mathbb{N}= \M\times\R$ with a terminal manifold of dimension
$(m+1)-k$ defined by the set of equations $\{S_k(\x,\ti)=0\}$
where $k\leq m$, $\x\in\M$, and $\ti\in [\ti_a,\ti_b]\subseteq
\R$. Let
\begin{equation}
  I(x)=\int_{\ti_a}^{\ti_b}F(\ti,x,\dot{x})\,d\ti\nonumber
\end{equation}
be the functional to be analyzed with $x:\R\rightarrow\M$. The
extrema of $I(x)$ solve the variational problem for
point-to-boundary paths. I will refer to the solutions as
\emph{critical paths} and denote them by $x_{cr}(\ti)$. For the
case of $\M=\R^n$, the variational problem is solved by the usual
Euler equations supplemented by `transversality' conditions (see
for example chapter six of \cite{SA}).

There are two limiting cases of interest. When the terminal
manifold in $\mathbb{N}$ coincides with the boundary in $\M$, then
$k=1$ and the transversality conditions reduce to
\begin{equation}\label{transversality}
  F(\ti_b,x(\ti_b),\dot{x}(\ti_b))
  =-\nu\nabla S(x(\ti_b))\cdot\dot{\Bold{x}}(\ti_b)
\end{equation}
where $\nu\neq 0$ is a constant. Thus critical paths, which
satisfy the Euler equations and this constraint, solve the
variational problem for paths with fixed initial point that
intersect a boundary. For free motion, (\ref{transversality})
implies the critical paths intersect the boundary transversally.

The other case of interest is when the manifold in the
dependent-independent space is ``horizontal'', i.e. $x(\ti_b)$ is
fixed and the terminal manifold is a line along the $\ti$
direction. The terminal manifold is determined by $k=m$ equations
and the transversality conditions yield
\begin{equation}\label{constant energy}
 F(\ti_b,x(\ti_b),\dot{x}(\ti_b))
  =\nabla_{\dot{\mathbf{e}}} F(\ti_b,x(\ti_b),
  \dot{x}(\ti_b))\cdot\dot{\Bold{x}}(\ti_b)
\end{equation}
where $\dot{\mathbf{e}}$ is a unit vector in the $\dot{\Bold{\x}}$
direction. If, in particular, $F=L+E$ where $L$ is the Lagrangian
of an isolated physical system and $E$ is a constant, then this is
just the fixed energy constraint $(\partial L/\partial
\dot{x}^i)\dot{x}^i-L=E$. Consequently, the variational problem in
this case is solved by paths with both end-points fixed that have
constant energy. This gives some insight into the connection
between fixed-energy path integrals and the elliptic Dirichlet
problem.

 To summarize, the presence of a boundary in the variational
 problem requires the introduction of a dependent variable $\tr$.
 For paths ending anywhere on a boundary, the critical paths satisfy a
 transversality condition. For paths ending at a fixed point within a bounded region,
 the critical paths satisfy a constant energy constraint. Inasmuch as the
 critical paths dominate in a path integral, the variational
 results motivate the use of $\tr$ in a path integral construction.
 Moreover, the evaluation of a path integral often requires explicit
 expressions for the critical paths: therefore, the
 transversality and fixed energy constraints can be expected to play a role in
 actual calculations.

\appendix

\bibliography{JohnBib1}

\begin{thebibliography}{99}
\bibitem{DY/YU} E.B. Dynkin and A.A. Yushkevich, \emph{Markov
Processes} (Plenum Press, New York, 1969).
\bibitem{FR} M. Freidlin, \emph{Functional Integration and
Partial Differential Equations} (Princeton University Press,
Princton, New Jersey, 1985).
\bibitem{FRI1}A. Friedman, \emph{Stochastic Differential
Equations and Applications, Vols. 1 and 2} (Academic Press, New
York, 1969).
\bibitem{FE/HI}R.P. Feynman and A.R. Hibbs, \emph{Quantum
Mechanics and Path Integrals} (McGraw-Hill, New York, 1965).
\bibitem{CA/D-M} P. Cartier and C. DeWitt-Morette, ``A new
perspective on functional integration,'' J. Math. Phys.
\textbf{36}, 2237 (1995).
\bibitem{LA} J. Lachapelle, ``Path integral solution of the
Dirichlet problem,'' Ann. Phys. \textbf{254}(2), 397
(1997).
\bibitem{EL} K.D. Elworthy, \emph{Stochastic Differential
Equations on Manifolds} (Cambridge University Press, Cambridge,
1982).
\bibitem{CA/D-W2} P. Cartier and C. DeWitt-Morette, ``A rigorous
mathematical foundation of functional integration,'' in
\emph{Functional Integration: Basics and Applications}, edited by
C. DeWitt-Morette (Plenum Press, New York, 1997).
\bibitem{FE}R.P. Feynman, \emph{The Principle of Least Action in Quantum
Mechanics} (PhD thesis, Princeton University, 1942).
\bibitem{GR1} C. Grosche, ``$\delta'$-function perturbations and
Neumann boundary-conditions by path integration,'' DESY Preprint,
DESY 94-019.
\bibitem{GR2} C. Grosche, ``Path integrals for two- and
three-dimensional $\delta$-function perturbations,'' Ann. der
Physik \textbf{3}, 283 (1994).
\bibitem{GR3} C. Grosche, ``Boundary-conditions in path
integrals,'' DESY Preprint, DESY 95-032.
\bibitem{AB/MA} R. Abraham and J.E. Marsden, \emph{Foundations
of Mechanics} (Benjamin/Cummings, London, 1978).
\bibitem{SA} H. Sagan, \emph{Introduction to the Calculus of
Variations} (McGraw-Hill, New York, 1969).
\bibitem{CA2} P. Cartier, ``A course on determinants,'' in
\emph{Conformal Invariance and String Theory}, edited by P. Dita
and V. Georgescu (Academic Press, New York, 1989).
\bibitem{CA/BE/D-W/WU} P. Cartier, M. Berg, C. DeWitt-Morette, and
A. Wurm, ``Characterizing volume forms,'' IHES Preprint,
IHEW-P/00/xx.
\bibitem{YO/D-W} A. Young and C. DeWitt-Morette, ``Time
substitutions in stochastic processes as a tool in path
integration,'' Ann. Phys. \textbf{169}(1), 140 (1986).
\bibitem{DU/KL} I.H. Duru and H. Kleinert, ``Quantum mechanics of
H-atom from path integrals,'' Fortschr. Phys. \textbf{30}, 401
(1982).
\bibitem{D-W/MA/NE} C. DeWitt-Morette, A. Maheshwari, and B.
Nelson, ``Path integration in non-relativistic quantum
mechanics,'' Physics Reports \textbf{50}(5), (1979).
\bibitem{LA/D-M} M.G.G. Laidlaw and C. DeWitt-Morette, ``Feynman
functional integrals for systems of indistinguishable particles,''
Phys. Rev. D \textbf{3}, 1375 (1971).
\bibitem{DO} J.S. Dowker, ``Quantum mechanics and field theory on
multiply connected and on homogeneous spaces,'' J. Phys. A
\textbf{5}, 936 (1972).
\bibitem{FRI2} A. Friedman, \emph{Partial Differential Equations
of Parabolic Type} (Prentice-Hall, New Jersey, 1964).
\bibitem{MI} C. Miranda, \emph{Partial Differnetial Equation of
Elliptic Type} (Springer-Verlag, New York, 1970).
\bibitem{GI/TR} D. Gilbarg and N.S. Trudinger, \emph{Elliptic
Partial Differential Equations of Second Order} (Springer-Verlag,
Berlin, 1983).
\bibitem{GR} P.A. Griffiths, \emph{Exterior Differential Systems
and the Calculus of Variations} volume 25 of \textit{Progress in
Mathematics} (Birkhauser, Boston, 1983).
\bibitem{DU/XU} C.F. Dunkl and Y. Xu, \emph{Orthogonal Polynomials
of Several Variables} (Cambridge University Press, Cambridge, UK
2001).
\bibitem{GL/JA} J. Glimm and A. Jaffe, \emph{Quantum Physics}
(Springer-Verlag, New York, 1987.)
\end{thebibliography}
\bibliographystyle{unsrt}

\end{document}